\documentclass[journal]{IEEEtran}

\IEEEoverridecommandlockouts                              

\usepackage{amssymb}
\usepackage{amsmath}
\usepackage{bm}
\usepackage{mathrsfs}
\usepackage{comment} 
\usepackage{algorithmic,algorithm}
\usepackage{shuffle}
\usepackage{graphicx} 
\usepackage{caption}
\usepackage{lipsum}
\usepackage{color}
\usepackage{enumerate}
\usepackage{amsfonts}
\usepackage{subfigure} 
\usepackage{array}
\usepackage{enumerate}
\usepackage{setspace}
\usepackage{multirow}

\DeclareSymbolFont{symbolsC}{U}{pxsyc}{m}{n}
\DeclareMathSymbol{\coloneqq}{\mathrel}{symbolsC}{"42}

\begin{document}

\title{\LARGE \bf
Privacy-Preserving Supervisory Control of Discrete-Event Systems via Co-Synthesis of Edit Function and Supervisor for Opacity Enforcement and Requirement Satisfaction} 
\author{Ruochen Tai, Liyong Lin, Yuting Zhu and Rong Su

%
%
%
\thanks{The research of the project was supported by Ministry of Education, Singapore, under grant AcRF TIER 1-2018-T1-001-245 (RG 91/18).

The authors are affliated with Nanyang Technological University, Singapore. (Email: ruochen001@e.ntu.edu.sg; liyong.lin@ntu.edu.sg; yuting002@e.ntu.edu.sg; rsu@ntu.edu.sg).
(\emph{Corresponding author: Liyong Lin})}
}
\maketitle

\begin{abstract}
This paper investigates the problem of co-synthesis of edit function and supervisor for opacity enforcement in the supervisory control of discrete-event systems (DES), assuming the presence of an external (passive) intruder, where the following goals need to be achieved: 1) the external intruder should never infer the system secret, i.e., the system is opaque, and never be sure about the existence of the edit function, i.e., the edit function remains covert; 2) the controlled plant behaviors should satisfy some safety and nonblockingness requirements, in the presence of the edit function. We focus on the class of edit functions that satisfy the following properties: 1) the observation capability of the edit function in general can be different from those of the supervisor and the intruder; 2) the edit function can implement insertion, deletion, and replacement operations; 3) the edit function performs bounded edit operations, i.e., the length of each string output of the edit function is upper bounded by a given constant. We propose an approach to solve this co-synthesis problem by modeling it as a distributed supervisor synthesis problem in the Ramadge-Wonham supervisory control framework. By taking the special structure of this distributed supervisor synthesis problem into consideration and to improve the possibility of finding a non-empty distributed supervisor, we propose two novel synthesis heuristics that incrementally synthesize the supervisor and the edit function. The effectiveness of our approach is illustrated on an example in the enforcement of the location privacy.  
\end{abstract}


\section{Introduction}
\label{intro}

With the development of Internet and mobile devices, we are now in an era of information explosion and big data, which not only brings about tremendous advantages, for example, better decision-making  capability, increased productivity, and improved agility, but also results in many challenges when implementing big data analytics initiatives. Storing big data, particularly sensitive data, can make the system a more attractive target for cyber attackers, one kind of which aims to infer the secret states of the system. To defend against such attackers and enforce security property, opacity came into being and was first introduced in computer science to analyze cryptographic protocols \cite{Mazare2004Opacity}.

In the context of DES, opacity is an attribute that expresses system security in a general language based theoretical framework. Its parameters are a predicate, given as a subset of runs of the system, and an observation function, from the set of runs into a set of observables \cite{Berard2015QuantifyOpacity}. If the secret cannot be inferred through observation by the external intruder, then the information is opaque.
Depending on the type of behaviour that is considered secret, two families of opacity properties are usually considered: state-based opacity \cite{Bryans2005SBO,Saboori2007SBO} and language-based opacity \cite{Badouel2007LBO,Dubreil2008LBO}, where the difference is that, for the state-based opacity, one is given a subset of secret states, while for the language-based opacity, one is given a subset of secret strings. Then the question is whether there exists some information sequences such that the observations generated by these sequences enable an external intruder to infer that the system state has transited to a secret state or a secret string has been executed by the system. For the state-based opacity, five kinds of derived opacity notions are mostly studied: 1) current-state opacity \cite{Bryans2005SBO,Saboori2007SBO}; 2) initial-state opacity \cite{Bryans2005SBO}; 3) initial-and-final-state opacity \cite{Wu2013ComparativeOpacity}; 4) $K$-step opacity \cite{Saboori2007SBO}; 5) infinite-step opacity \cite{Saboori2009InfiniteSBO}.

In the DES community, a substantial amount of studies have been focusing on opacity, including verification and enforcement. In this work, we shall focus on the opacity enforcement. For the opacity verification, \cite{Lin2011OpacityLBO} provides algorithms for checking strong and weak language-based opacity and  verification algorithms for state-based opacity properties are proposed in \cite{Saboori2007SBO}, \cite{Wu2013ComparativeOpacity}, \cite{Saboori2008VerifyISBO} - \cite{Yin2017InfKOpacity}. 
By modeling the system as a Petri net, \cite{Tong2017PetriOpacity,Tong2017PetriDecidability} address the verification of state-based opacity, where the decidability issue is considered in \cite{Tong2017PetriDecidability}. In addition, by modeling the system as a probabilistic finite state automaton, \cite{Yin2019PetriDecidability}-\cite{Keroglou2018ProbabilisticOpacity} investigate the verification of opacity in the context of stochastic DES, where the violation of opacity is characterized by the probability, not a binary value (0 or 1) as in the case of a non-stochastic DES. Furthermore, the investigation on the verification of opacity has also been extended to  networked DES recently in \cite{YinOpacityNetwork,Yang2021NetworkedOpacity}, where the communication delays and losses in the observation channel and the control channel have been taken into consideration. Readers could refer to \cite{Jacob2016OverviewOpacity,Lafortune2018OverviewOpacity} for a more comprehensive literature review.

\begin{figure}[htbp]
\centering
\subfigure[]{
\begin{minipage}[t]{0.3\linewidth}
\centering
\includegraphics[height=0.88in]{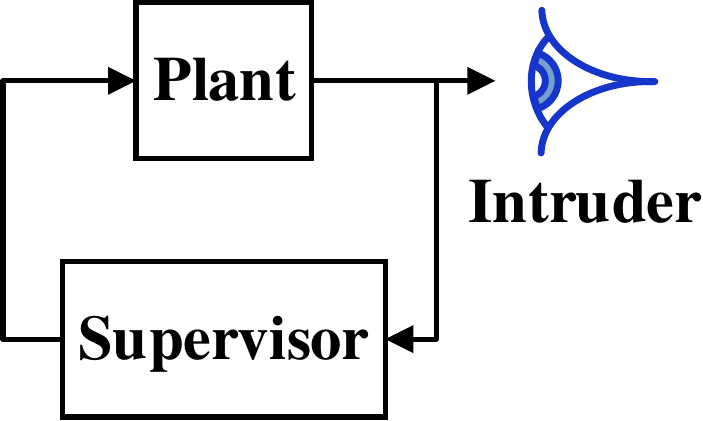}
\end{minipage}
}%

\subfigure[]{
\begin{minipage}[t]{0.3\linewidth}
\centering
\includegraphics[height=0.46in]{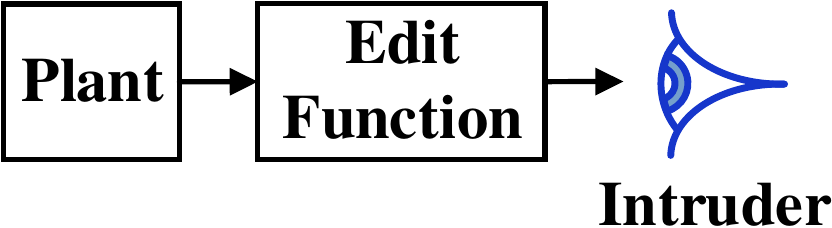}
\end{minipage}
}%
\subfigure[]{
\begin{minipage}[t]{0.85\linewidth}
\centering
\includegraphics[height=0.46in]{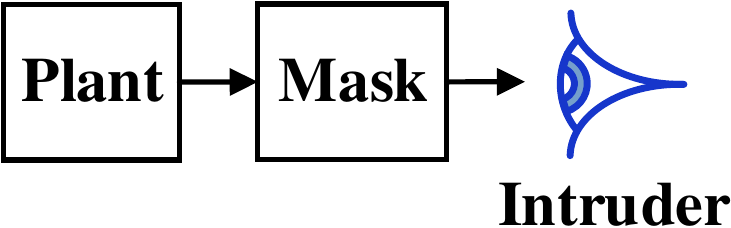}
\end{minipage}
}%

\centering
\caption{Three approaches for opacity enforcement: (a) Supervisory control. (b) Edit function. (c) Mask.}
\label{fig:Two approaches for opacity enforcement}
\end{figure}

For opacity enforcement, there are typically three approaches, which are shown in Fig. \ref{fig:Two approaches for opacity enforcement}: 1) Supervisory control, which restricts the system’s behavior such that the secret can be preserved; 2) Edit function, which modifies the information flow such that the external intruder cannot infer the system secret; 3) Mask, which turns on/off
the associated sensors to enforce the opacity.  

In \cite{Takai2009SCT}-\cite{Dubreil2008JDESOpacitySCT}, the technique of adopting supervisory control for opacity enforcement is investigated, where  maximally permissive controllers are synthesized. Specifically, \cite{Darondeau2014JDESSCT} specifies the finite transition systems as modal transition systems to ensure opacity of a secret predicate on all labeled transition systems. To mitigate the complexity of the synthesis
procedure, \cite{Hou2019OpacityNetwork} proposes abstraction-based synthesis of opacity-enforcing controllers by using alternating simulation relations for labeled transition systems. 

The topic of opacity enforcement by using edit functions is investigated in \cite{Falcone2015OpacityRuntime}, \cite{Wu2014AutoInsert}-\cite{Falcone2013CDCRuntime}, all of which assume that the edit function and the external intruder have the same observation capability and could observe all the observable events fired by the plant. \cite{Wu2014AutoInsert} considers the problem of enforcing current-state opacity and language-based opacity by using insertion functions. \cite{Barcelos2018WODESShuffle} deals with current-state opacity and proposes an enforcer to change the order of observations in the event occurrences. \cite{Ji2018AutoPublicInsert}-\cite{Wu2015CDCPublicInsert} study the problem of enforcing current-state opacity under the assumption that the intruder either knows or does not know the structure of the insertion function. In addition, deletion functions are considered in \cite{Ji2017CDCPublicInsert}, which is also extended to nondeterministic insertion and deletion functions in \cite{Ji2019TACNDPublicEdit}. To reduce the computational complexity, \cite{Mohajerani2018CDCAbstractions} proposes abstraction based methods to synthesize edit functions for current-state opacity enforcement and then \cite{Mohajerani2019TACAbstractions} extends the work in \cite{Mohajerani2018CDCAbstractions} by taking the synchronous composition into consideration under modular DES. \cite{Falcone2015OpacityRuntime,Falcone2013CDCRuntime} adopt runtime enforcer, which enforces opacity by using delays, to ensure $K$-step opacity.

For the techniques of adopting masks to enforce opacity, \cite{Cassez2012FMSDMask} designs masks to restrict the observable outputs of the system either in a static or dynamic way to ensure current-state opacity.
\cite{Yin2019TACMask} investigates the problem of synthesizing dynamic masks that preserve infinite-step opacity.
\cite{Zhang2015TASEMask} studies the problem of maximum information release while ensuring (weak or strong) language-based opacity.

As we have described above, lots of fruitful works have been dedicated to opacity enforcement of DES. However, existing research only considers either synthesis of supervisor to restrict the system behavior or synthesis of edit function or mask to ensure that the information flow is opaque when the system behavior is not restricted. In reality, it is more likely that the system behavior is restricted meanwhile we need to enforce opacity w.r.t such restricted behavior. Thus, the following privacy-preserving control problem needs to be solved. On one hand, the system needs to fulfill some specific requirement, which might not satisfy the opacity property, by adopting supervisory control, and on the other hand, we expect that the information sequences generated by the system would not expose the system secret to the external intruder by adopting the edit functions or masks. In this work, we choose to adopt edit functions. To achieve the goals in this privacy-preserving control problem, the edit function and the supervisor ought to cooperate to control the system and confuse the intruder. However, the difficulty is that what the supervisor observes is the information sequence altered by the edit function, which is originally used to deceive the intruder but it might also confuse the supervisor. Thus, the edit function and the supervisor should be designed carefully enough such that only the intruder would be confused and the supervisor could still issue the appropriate control commands under the altered information sequences.

In this work, besides the opacity enforcement that should be guaranteed in the above-mentioned privacy-preserving control issue, we also take the covertness into consideration when we synthesize the edit function. In the previous works, it is usually assumed that the external intruder has the full knowledge of the plant as its prior knowledge, based on which it could infer the system secret. In this paper, we consider a more powerful intruder that could not only infer the secret but also could discover the existence of edit function, since the intruder can compare its online observations with its prior knowledge to determine whether information inconsistency has happened. We assume that once the intruder detects such inconsistency, the existence of the edit function is exposed to the intruder, i.e., the edit function is not covert. Our goal is that the synthesized edit function should always remain covert to the intruder, making it as ambiguous as possible for the intruder, which imposes more challenges when we synthesize the edit function and the supervisor, since now the feasible edit operations initiated by the edit function should not only ensure the opacity but also cannot expose its own existence.

In this work, we shall study a privacy-preserving control issue, by focusing on the problem of co-synthesizing the edit function and the supervisor for opacity enforcement in the supervisory control of DES. To the best of our knowledge, this is the first time when  such a synthesis problem is investigated in the context of DES. The contributions of this work are as follows:
\begin{enumerate}[1.]
\setlength{\itemsep}{3pt}
\setlength{\parsep}{0pt}
\setlength{\parskip}{0pt}
    \item We consider the privacy-preserving supervisory control issue by addressing the problem of co-synthesis of the edit function and the supervisor, which is more in line with the need for the resilient control of a closed-loop system. In this work, we adopt a general setup for this privacy-preserving control problem, where the observation capabilities of the edit function, the supervisor, and the intruder could be different. This general setup has never been considered in previous works on opacity enforcement with edit functions. In addition, we also consider the covertness enforcement for the edit function, so the external intruder is never sure whether there exists an edit function.
    \item By formulating the system components as finite state automata, the problem of co-synthesizing the edit function and the supervisor for opacity enforcement is addressed. The solution methodology proposed in this work is to model the co-synthesis problem as a distributed supervisor synthesis problem in the Ramadge-Wonham supervisory control framework.
    \item To solve the co-synthesis problem, which has been modelled as a distributed supervisor synthesis problem, we propose two incremental synthesis heuristics that exploit the structure of the distributed control architecture arising from the co-synthesis of the supervisor and the edit function. 
    Different from the existing incremental synthesis approaches, which attempt to synthesize a nonblocking local supervisor at each step and can immediately result in an empty solution for our problem, our approach avoids this pitfall and attempts to synthesize a local supervisor to ensure the marker-reachability first, which can thus increase the possibility of generating a feasible solution for the distributed supervisor synthesis problem studied in this work. 
\end{enumerate}

This paper is organized as follows. In Section \ref{sec:Preliminaries}, we provide some basic notions which are needed in this work. In Section \ref{sec:Component Models of DES under Edit Function and Supervisor}, we introduce the component models that help us to model the co-synthesis problem as a distributed supervisor synthesis problem. Section \ref{sec:Co-Synthesis of Edit Function and Supervisor for Opacity Enforcement} proposes a method to synthesize the edit function and the supervisor for opacity enforcement. An example is given to show the effectiveness of the proposed method in Section \ref{sec:example}. Finally, conclusions are drawn in Section \ref{sec:conclusions}.

\section{Preliminaries}
\label{sec:Preliminaries}
Given a finite alphabet $\Sigma$, let $\Sigma^{*}$ be the free monoid over $\Sigma$ with the empty string $\varepsilon$ being the unit element and the string concatenation being the monoid operation. For a string $s$, $|s|$ is defined as the length of $s$. Given two strings $s, t \in \Sigma^{*}$, we say $s$ is a prefix substring of $t$, written as $s \leq t$, if there exists $u \in \Sigma^{*}$ such that $su = t$, where $su$ denotes the concatenation of $s$ and $u$. A language $L \subseteq \Sigma^{*}$ is a set of strings. The prefix closure of $L$ is defined as $\overline{L} = \{u \in \Sigma^{*} \mid (\exists v \in L) \, u\leq v\}$. 
The event set $\Sigma$ is partitioned into $\Sigma = \Sigma_{c} \dot{\cup} \Sigma_{uc} = \Sigma_{o} \dot{\cup} \Sigma_{uo}$, where $\Sigma_{c}$ (respectively, $\Sigma_{o}$) and $\Sigma_{uc}$ (respectively, $\Sigma_{uo}$) are defined as the sets of controllable (respectively, observable) and uncontrollable (respectively, unobservable) events, respectively.  As usual, $P_{o}: \Sigma^{*} \rightarrow \Sigma_{o}^{*}$ is the natural projection defined such that
\begin{enumerate}[(1)]
\setlength{\itemsep}{3pt}
\setlength{\parsep}{0pt}
\setlength{\parskip}{0pt}
\item $P_{o}(\varepsilon) = \varepsilon$,
\item $(\forall \sigma \in \Sigma) \, P_{o}(\sigma)=
\left\{
\begin{array}{rcl}
\sigma       &      & {\sigma \in \Sigma_{o},}\\
\varepsilon  &      & {\rm otherwise,}
\end{array} \right.$
\item $(\forall s \in \Sigma^*, \sigma \in \Sigma) \, P_{o}(s\sigma) = P_{o}(s)P_{o}(\sigma)$.
\end{enumerate}

A finite state automaton $G$ over $\Sigma$ is given by a 5-tuple $(Q, \Sigma, \xi, q_{0}, Q_{m})$, where $Q$ is the state set, $\xi: Q \times \Sigma \rightarrow Q$ is the (partial) transition function, $q_{0} \in Q$ is the initial state, and $Q_{m}$ is the set of marker states. 
We write $\xi(q, \sigma)!$ to mean that $\xi(q, \sigma)$ is defined and also view $\xi \subseteq Q \times \Sigma \times Q$ as a relation. $En_{G}(q) = \{\sigma \in \Sigma|\xi(q, \sigma)!\}$.
$\xi$ is also extended to the (partial) transition function $\xi: Q \times \Sigma^{*} \rightarrow Q$ and the transition function $\xi: 2^{Q} \times \Sigma \rightarrow 2^{Q}$ \cite{wonham2015supervisory}, where the later is defined as follows: for any $Q' \subseteq Q$ and any $\sigma \in \Sigma$, $\xi(Q', \sigma) = \{q' \in Q|(\exists q \in Q')q' = \xi(q, \sigma)\}$. 
Let $L(G)$ and $L_{m}(G)$ denote the closed-behavior and the marked behavior of $G$~\cite{wonham2015supervisory}, respectively. When $Q_{m} = Q$, we shall also write $G = (Q, \Sigma, \xi, q_{0})$ for simplicity. 
The ``unobservable reach''\cite{wonham2015supervisory} of the state $q \in Q$ under the subset of events $\Sigma' \subseteq \Sigma$ is given by $UR_{G, \Sigma - \Sigma'}(q) := \{q' \in Q|[\exists s \in (\Sigma - \Sigma')^{*}] \, q' = \xi(q,s)\}$.
We define  $P_{\Sigma'}(G)$ to be the finite state automaton $(2^{Q}, \Sigma, \delta, UR_{G, \Sigma - \Sigma'}(q_{0}))$ over $\Sigma$, where the unobservable reach $UR_{G, \Sigma - \Sigma'}(q_{0})$ of $q_{0}$ is the initial state, and the (partial) transition function $\delta: 2^{Q} \times \Sigma \rightarrow 2^{Q}$ is defined as follows:
\begin{enumerate}[(1)]
    \item For any $\varnothing \neq Q' \subseteq Q$ and any $\sigma \in \Sigma'$, $\delta(Q', \sigma) = UR_{G, \Sigma - \Sigma'}(\xi(Q', \sigma))$, where
    \[
    UR_{G, \Sigma - \Sigma'}(Q'') = \bigcup\limits_{q \in Q''}UR_{G, \Sigma - \Sigma'}(q)
    \]
    for any $Q'' \subseteq Q$;
    \item For any $\varnothing \neq Q' \subseteq Q$ and any $\sigma \in \Sigma - \Sigma'$, $\delta(Q', \sigma) = Q'$.
\end{enumerate}
We here remark that $P_{\Sigma'}(G)$ is over $\Sigma$, instead of $\Sigma'$, and there is no transition defined at the state $\varnothing \in 2^{Q}$.

A finite state automaton $G = (Q, \Sigma, \xi, q_{0}, Q_{m})$ is said to be nonblocking if every reachable state in $G$ can reach some marker state in $Q_{m}$~\cite{wonham2015supervisory}, and marker-reachable if some marker state in $Q_m$ is reachable. As usual, for any two finite state automata $G_{1} = (Q_{1}, \Sigma_{1}, \xi_{1}, q_{1,0}, Q_{1,m})$ and $G_{2} = (Q_{2}, \Sigma_{2}, \xi_{2}, q_{2,0}, Q_{2,m})$, where $En_{G_{1}}(q) = \{\sigma|\xi_{1}(q, \sigma)!\}$ and $En_{G_{2}}(q) = \{\sigma|\xi_{2}(q, \sigma)!\}$, their synchronous product \cite{CassandrasDES2008} is denoted as $G_{1}||G_{2} := (Q_{1} \times Q_{2}, \Sigma_{1} \cup \Sigma_{2}, \zeta, (q_{1,0}, q_{2,0}), Q_{1,m} \times Q_{2,m})$, where the (partial) transition function $\zeta$ is defined as follows: for any $(q_{1}, q_{2}) \in Q_{1} \times Q_{2}$ and $\sigma \in \Sigma$:
\[
\begin{aligned}
& \zeta((q_{1}, q_{2}), \sigma) := \\ & \left\{
\begin{array}{lcl}
(\xi_{1}(q_{1}, \sigma), \xi_{2}(q_{2}, \sigma))  &      & {\rm if} \, {\sigma \in En_{G_{1}}(q_{1}) \cap En_{G_{2}}(q_{2}),}\\
(\xi_{1}(q_{1}, \sigma), q_{2})       &      & {\rm if} \, {\sigma \in En_{G_{1}}(q_{1}) \backslash \Sigma_{2},}\\
(q_{1}, \xi_{2}(q_{2}, \sigma))       &      & {\rm if} \, {\sigma \in En_{G_{2}}(q_{2}) \backslash \Sigma_{1},}\\
{\rm not \, defined}  &      & {\rm otherwise.}
\end{array} \right.
\end{aligned}
\]
\textbf{Notation.} Let $\mathbb{Z}$ denote the set of integers, $\mathbb{N}$ the set of nonnegative integers, and $\mathbb{N}^{+}$ the set of positive integers. Let $\Gamma = 2^{\Sigma_c}-\{\varnothing\}$ denote the set of all the possible control commands, deviating from the standard definition of $\Gamma$, where each control command only contains the controllable events that it will enable. It is assumed that uncontrollable events could be executed independently of a control command.
For an alphabet $\Sigma$, we use $\Sigma^{\#}$ to denote a copy of $\Sigma$ with superscript ``$\#$'' attached to each element in $\Sigma$. Intuitively speaking, ``$\sigma^{\#}$'' denotes the message edited by the edit function and the specific meanings of the relabelled events will be introduced later in Section \ref{sec:Component Models of DES under Edit Function and Supervisor}.

\section{Component Models with Edit Function and Supervisor}
\label{sec:Component Models of DES under Edit Function and Supervisor}

\begin{figure*}[htbp]
\begin{center}
\includegraphics[height=6.7cm]{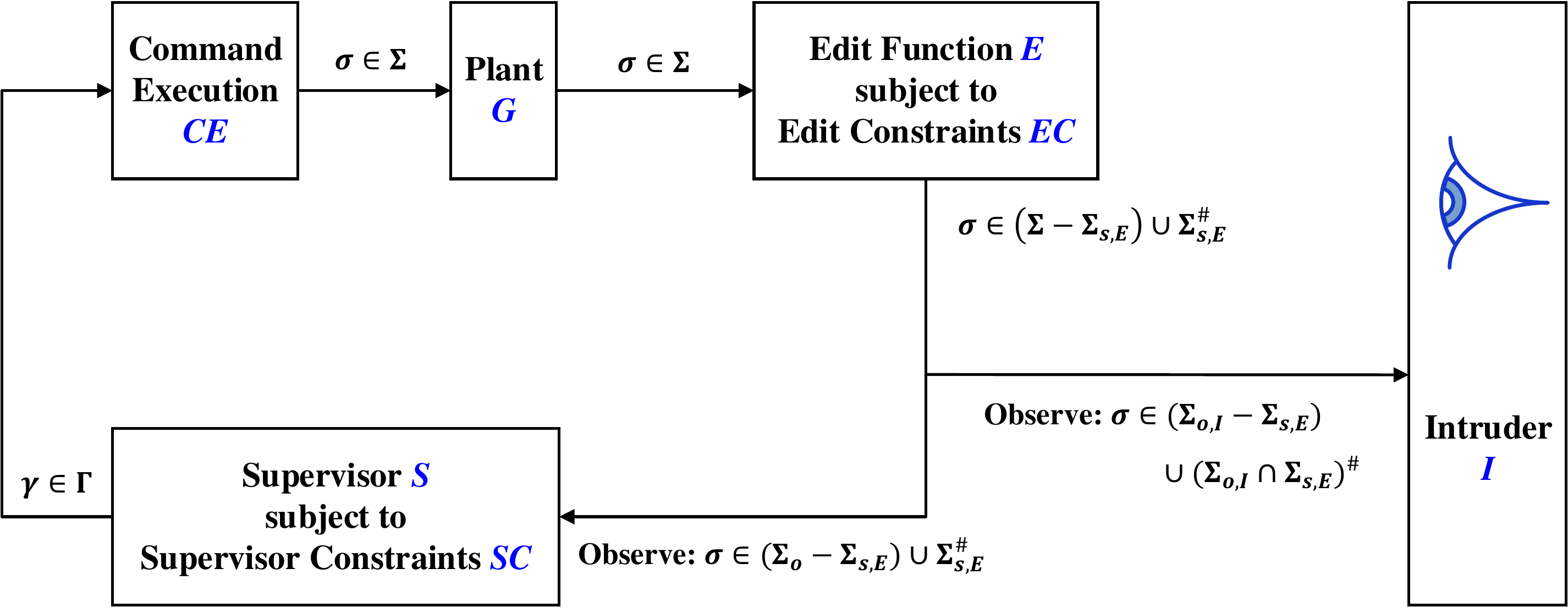}   
\caption{Control architecture under edit function and supervisor for opacity enforcement}
\label{fig:Supervisory control architecture under edit function}
\end{center}        
\end{figure*}

The architecture of the privacy-preserving supervisory control system with an edit function for opacity enforcement is illustrated in Fig. \ref{fig:Supervisory control architecture under edit function}, where the components are listed as follows:
\begin{itemize}
\setlength{\itemsep}{3pt}
\setlength{\parsep}{0pt}
\setlength{\parskip}{0pt}
    \item Edit function $E$ (subject to edit constraints $EC$).
    \item Supervisor $S$ (subject to supervisor constraints $SC$).
    \item Command execution component $CE$. 
    \item Plant $G$.
    \item Intruder $I$.
\end{itemize}

In the following subsections, we shall explain how we model the above-mentioned five components. 

\subsection{Edit function}
\label{subsec:Edit function}
The set of observable events for the edit function is denoted as $\Sigma_{o,E} \subseteq \Sigma_{o}$, where $\Sigma_o$ denotes the set of observable events for the supervisor. The set of editable events for the edit function is denoted as $\Sigma_{s,E} \subseteq \Sigma_{o,E}$, that is, the edit function could only delete, insert, and replace events in $\Sigma_{s,E}$.  

The basic assumptions of the edit function in this work are given as follows:
\begin{itemize}
    \item The edit function could implement insertion, deletion, and  replacement operations.
    \item In Fig. \ref{fig:Supervisory control architecture under edit function}, any event $\sigma \in \Sigma_{o,E}$ fired by the plant $G$ will be firstly observed by the edit function, then the output of the edit function, if observable to the intruder, would be eavesdropped by the intruder.
    \item The edit function  carries out edit operation each time when it observes some event in $\Sigma_{o,E}$. Each time when the edit function observes one event in $\Sigma_{o,E}$, the number of events that it can simultaneously send to the supervisor is bounded by $U$, i.e., we consider bounded edit function.
    \item The edit action initiated by the edit function is instantaneous.
\end{itemize}

Next, we shall introduce two models that will be used in this work: 1) edit constraints; 2) edit function, where the former one serves as a ``template'' to describe the capabilities of the edit function and the latter one is the edit function that we aim to synthesize.

\begin{figure}[htbp]
\begin{center}
\includegraphics[height=3.4cm]{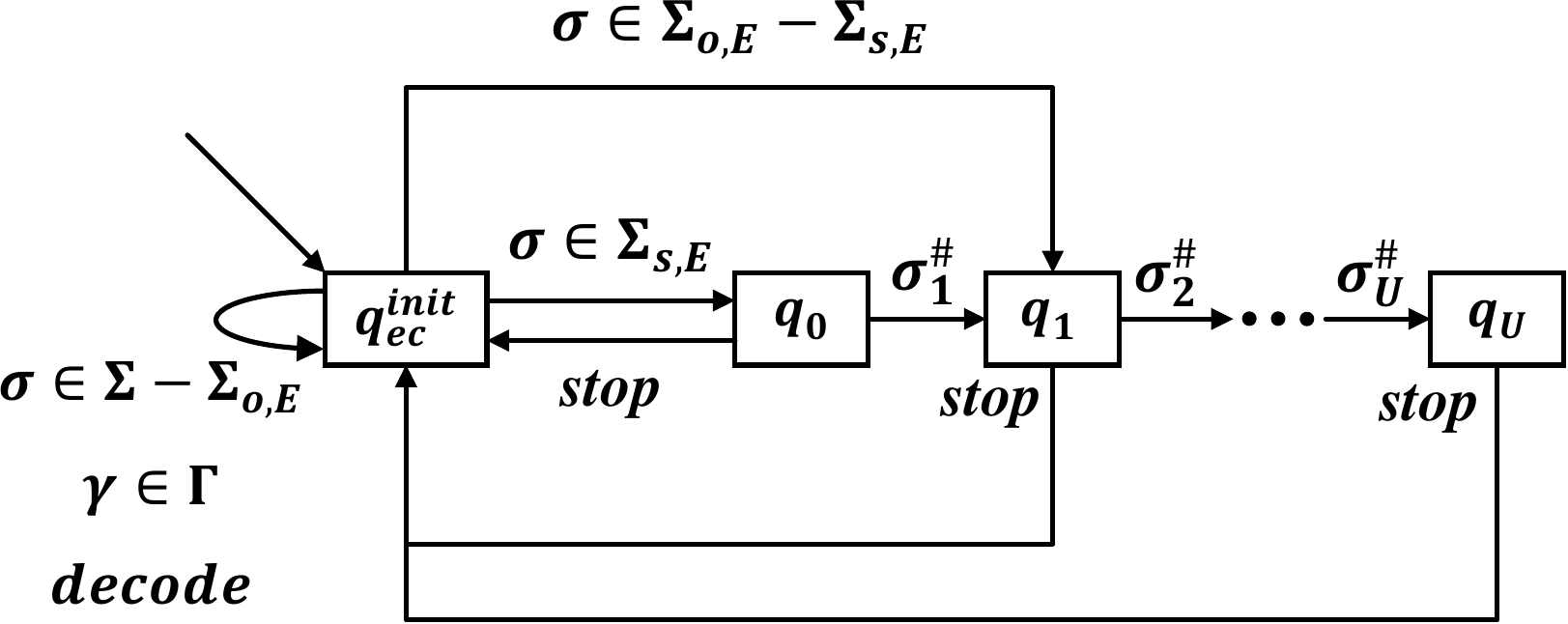}   
\caption{The (schematic) model for edit constraints ($\sigma_{1}, \dots, \sigma_{U} \in \Sigma_{s,E}$)}
\label{fig:The (schematic) model for editor constraints}
\end{center}        
\end{figure}

\textbf{Edit Constraints:} The edit constraints is modeled as a finite state automaton $EC$, which is shown in Fig. \ref{fig:The (schematic) model for editor constraints}.  
\[
EC = (Q_{ec}, \Sigma_{ec}, \xi_{ec}, q_{ec}^{init}, Q_{ec,m})
\]
\begin{itemize}
\setlength{\itemsep}{3pt}
\setlength{\parsep}{0pt}
\setlength{\parskip}{0pt}
    \item $Q_{ec} = \{q_{n}|n \in [0:U]\} \cup \{q_{ec}^{init}\}$
    \item $\Sigma_{ec} = \Sigma \cup \Sigma_{s,E}^{\#} \cup \Gamma \cup \{stop, decode\}$
    \item $\xi_{ec}: Q_{ec} \times \Sigma_{ec} \rightarrow Q_{ec}$
    \item $Q_{ec,m} = \{q_{ec}^{init}\}$
\end{itemize}
The (partial) transition function $\xi_{ec}$ is defined as follows:
\begin{enumerate}[1.]
\setlength{\itemsep}{3pt}
\setlength{\parsep}{0pt}
\setlength{\parskip}{0pt}
    \item For any $\sigma \in (\Sigma - \Sigma_{o,E}) \cup \Gamma \cup \{decode\}$, $\xi_{ec}(q_{ec}^{init}, \sigma) = q_{ec}^{init}$.
    \item For any $\sigma \in \Sigma_{s,E}$, $\xi_{ec}(q_{ec}^{init}, \sigma) = q_{0}$.
    \item For any $\sigma \in \Sigma_{o,E} - \Sigma_{s,E}$, $\xi_{ec}(q_{ec}^{init}, \sigma) = q_{1}$.
    \item For any $n \in [0:U-1]$ and any $\sigma \in \Sigma_{s,E}$, $\xi_{ec}(q_{n}, \sigma^{\#}) = q_{n+1}$.
    \item For any $n \in [0:U]$, $\xi_{ec}(q_{n}, stop) = q_{ec}^{init}$.
\end{enumerate}

Next, we shall present some explanations for the model $EC$. In the state set $Q_{ec}$,
\begin{itemize}
    \item $q_{ec}^{init}$ is the initial state. It is a state denoting that 1) no edit operation has been conducted, or 2) the edit function has not observed any event $\sigma \in \Sigma_{o,E}$ since the end of the last edit operation.
    \item $q_{n}(n \in [0:U])$ is a state denoting that the edit function has sent $n$ events since it observes some event \footnote{In this work, when the edit function observes some $\sigma \in \Sigma_{o,E} - \Sigma_{s,E}$, since it cannot edit $\sigma$ and would just let $\sigma$ pass, we shall count such event in the output sent by the edit function. Thus, the observation of $\sigma \in \Sigma_{o,E} - \Sigma_{s,E}$ at the edit function implies that the edit function sends $\sigma$.}. Specifically, at the state $q_{0}$, the edit function could either delete the observed event or replace the observed event with any editable event. At the state $q_{n}(n \in [1:U-1])$, the edit function could insert any editable event or end the current round of edit operation. At the state $q_{U}$, the edit function must end the current round of edit operation and cannot insert editable events anymore since the upper bound of the output is $U$.
\end{itemize}

In the event set $\Sigma_{ec}=\Sigma \cup \Sigma_{s,E}^{\#} \cup \Gamma \cup \{stop, decode\}$, $\sigma \in \Sigma$ denotes the firing of $\sigma$ by plant $G$, any $\sigma^{\#} \in \Sigma_{s,E}^{\#}$ denotes the event of sending an editable event $\sigma \in \Sigma_{s,E}$ by the edit function, and the event $stop$ denotes the end of current round of edit operation, which can be controlled and observed by the edit function. Any element in $\Gamma \cup \{decode\}$ denotes an event happening in other three components: supervisor, command execution component and intruder. Intuitively speaking, any element in $\Gamma$ denotes a control command issued by the supervisor and the event $decode$ denotes that the secret state of plant $G$ has been inferred by the intruder; these will be introduced later in Section \ref{subsec:Supervisor}, \ref{subsec:Command execution}, and \ref{subsec:Intruder}. All of the events in $\Gamma \cup \{decode\}$ are assumed to be unobservable and uncontrollable to the edit function in this work.

For the (partial) transition function $\xi_{ec}$,
\begin{itemize}
\setlength{\itemsep}{2pt}
\setlength{\parsep}{0pt}
\setlength{\parskip}{0pt}
    \item Case 1 says that, at state $q_{ec}^{init}$, if any event $\sigma \in (\Sigma - \Sigma_{o,E}) \cup \Gamma \cup \{decode\}$ happens, the edit function will not carry out any edit operation since it cannot observe $\sigma$, and such event will lead to a self-loop.
    \item Case 2 says that, at state $q_{ec}^{init}$, after the edit function observes any event $\sigma \in \Sigma_{s,E}$, it would transit to state $q_{0}$, at which it could either delete $\sigma$ or replace $\sigma$ with any editable event in $\Sigma_{s,E}$.
    \item Case 3 says that, at state $q_{ec}^{init}$, after the edit function observes any event $\sigma \in \Sigma_{o,E} - \Sigma_{s,E}$, it would transit to state $q_{1}$ and let $\sigma$ pass because it cannot edit $\sigma$. Since the number of events that the edit function can simultaneously send after it observes one event in $\Sigma_{o,E}$ is upper bounded by $U$, the edit function could still insert at most $U-1$ events in $\Sigma_{s,E}$ after observing $\sigma$.\footnote{In this work, we shall count $\sigma \in \Sigma_{o, E}-\Sigma_{s, E}$ in the events sent by the edit function. If readers prefer to not count $\sigma$ in the events sent by the edit function, then only minor modifications are needed. One possible way is to replace the transition $\xi_{ec}(q_{ec}^{init}, \sigma) = q_{1}$ with the transition $\xi_{ec}(q_{ec}^{init}, \sigma) = q_{0}$.}
    \item Case 4 says that at any state $q_{n}(n \in [0:U-1])$, the edit function could insert any editable event $\sigma \in \Sigma_{s,E}$. Since the number of the events that can be sent by the edit function after observing some event is upper bounded by $U$, at the state $q_{U}$, the edit function cannot insert any editable event.
    \item Case 5 says that at any state $q_{n}(n \in [0:U])$, the edit function could end the current round of the edit operation and transit back to the initial state $q_{ec}^{init}$ with the event $stop$.
\end{itemize}

Based on the model of  $EC$, it can be seen that, in the output event set of the edit function, all the events in $\Sigma_{s,E}$ are relabelled as the copies in $\Sigma_{s,E}^{\#}$ by attaching the superscript ``\#''. Based on the model of $EC$, we have $|Q_{ec}| = U+2$.

\textbf{Edit Function:} The edit function is modeled as a finite state automaton $E$.
\[
E = (Q_{e}, \Sigma_{e}, \xi_{e}, q_{e}^{init}, Q_{e,m})
\]
where $\Sigma_{e} = \Sigma_{ec} = \Sigma \cup \Sigma_{s,E}^{\#} \cup \Gamma \cup \{stop, decode\}$, that satisfies the following constraints:
\begin{itemize}
\setlength{\itemsep}{2pt}
\setlength{\parsep}{0pt}
\setlength{\parskip}{0pt}
    \item (E-controllability) For any state $q \in Q_{e}$ and any $\sigma \in \Sigma_{e,uc}: = \Sigma_{e} - \Sigma_{e,c} = \Sigma_{e} - (\Sigma_{s,E}^{\#} \cup \{stop\})$, $\xi_{e}(q,\sigma)!$.
    \item (E-observability) For any state $q \in Q_{e}$ and any $\sigma \in \Sigma_{e,uo} := \Sigma_{e} - \Sigma_{e,o} = \Sigma_{e} - (\Sigma_{o,E} \cup \Sigma_{s,E}^{\#} \cup \{stop\})$, if $\xi_{e}(q,\sigma)!$, then $\xi_{e}(q,\sigma) = q$.
\end{itemize}
E-controllability states that the edit function can only disable events in $\Sigma_{e,c} = \Sigma_{s,E}^{\#} \cup \{stop\}$. E-observability states that the edit function can only make a state change after observing an event in $\Sigma_{e,o} = \Sigma_{o,E} \cup \Sigma_{s,E}^{\#} \cup \{stop\}$. In this work, by construction, all the controllable events for the edit function are also observable to the edit function. In the following text, we shall refer to $(\Sigma_{e,c}, \Sigma_{e,o})$ as the edit function-control constraint.  

\subsection{Supervisor}
\label{subsec:Supervisor}
In this part, we shall introduce two models that will be used in this work: 1) supervisor constraints; 2) supervisor, where the former one serves as a ``template'' to describe the capabilities of the supervisor and the latter one is the supervisor that we aim to synthesize.

\begin{figure}[htbp]
\begin{center}
\includegraphics[height=3.8cm]{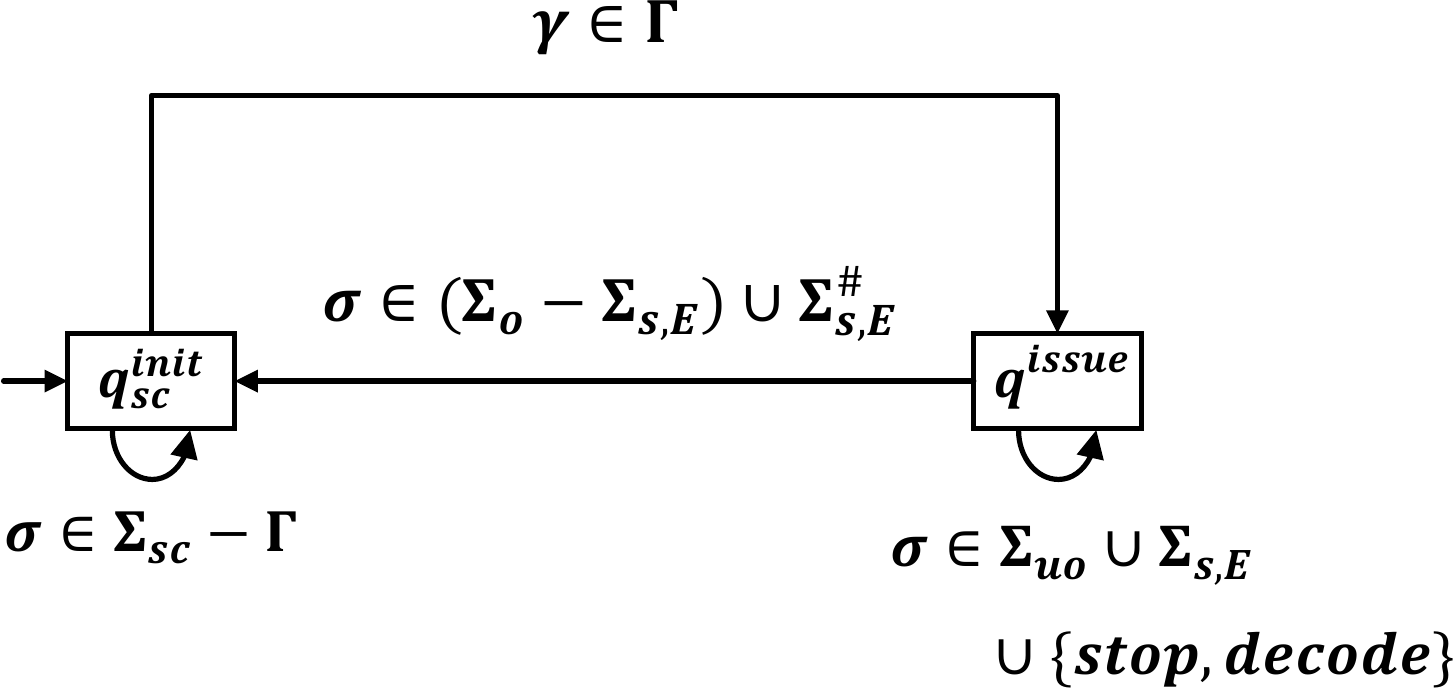}   
\caption{The (schematic) model for supervisor constraints}
\label{fig:The (schematic) model for supervisor constraints}
\end{center}        
\end{figure}

\textbf{Supervisor Constraints:} Firstly, due to the existence of the edit function, all the events in $\Sigma_{s,E}$ are relabelled in the output of the edit function, resulting in that the set of observed events by the supervisor is $(\Sigma_{o} - \Sigma_{s,E}) \cup \Sigma_{s,E}^{\#}$.
Then, the supervisor constraints is modeled as a finite state automaton $SC$, which is shown in Fig. \ref{fig:The (schematic) model for supervisor constraints}. 
Intuitively speaking, when the system initiates, the supervisor could issue an initial control command without observing any event in $(\Sigma_{o} - \Sigma_{s,E}) \cup \Sigma_{s,E}^{\#}$. Then, the supervisor could issue a new control command again only after it has observed at least one event in $(\Sigma_{o} - \Sigma_{s,E}) \cup \Sigma_{s,E}^{\#}$. In this work, we impose the natural assumption that the issued control command is observable to the supervisor.

\[
SC = (Q_{sc}, \Sigma_{sc}, \xi_{sc}, q_{sc}^{init})
\]
\begin{itemize}
\setlength{\itemsep}{3pt}
\setlength{\parsep}{0pt}
\setlength{\parskip}{0pt}
    \item $Q_{sc} = \{q_{sc}^{init}, q^{issue}\}$
    \item $\Sigma_{sc} = \Sigma \cup \Sigma_{s,E}^{\#} \cup \Gamma \cup \{stop, decode\}$
    \item $\xi_{sc}: Q_{sc} \times \Sigma_{sc} \rightarrow Q_{sc}$
\end{itemize}
The (partial) transition function $\xi_{sc}$ is defined as follows:
\begin{enumerate}[1.]
\setlength{\itemsep}{3pt}
\setlength{\parsep}{0pt}
\setlength{\parskip}{0pt}
    \item For any $\sigma \in \Sigma_{sc} - \Gamma = \Sigma \cup \Sigma_{s,E}^{\#} \cup \{stop, decode\}$, $\xi_{sc}(q_{sc}^{init}, \sigma) = q_{sc}^{init}$.
    \item For any $\gamma \in \Gamma$, $\xi_{sc}(q_{sc}^{init}, \gamma) = q^{issue}$.
    \item For any $\sigma \in (\Sigma_{o} - \Sigma_{s,E}) \cup \Sigma_{s,E}^{\#}$, $\xi_{sc}(q^{issue}, \sigma) = q_{sc}^{init}$.
    \item For any $\sigma \in \Sigma_{uo} \cup \Sigma_{s,E} \cup \{stop, decode\}$, $\xi_{sc}(q^{issue}, \sigma) = q^{issue}$.
\end{enumerate}

Next, we shall present some explanations for the model $SC$. In the state set $Q_{sc}$,
\begin{itemize}
\setlength{\itemsep}{3pt}
\setlength{\parsep}{0pt}
\setlength{\parskip}{0pt}
    \item $q_{sc}^{init}$ is the initial state. It is a state denoting that 1) the supervisor has not issued any control command since the system initiates, or 2) the supervisor has observed at least one event $\sigma \in (\Sigma_{o} - \Sigma_{s,E}) \cup \Sigma_{s,E}^{\#}$ since it issues the last control command. At state $q_{sc}^{init}$, the supervisor could choose to either wait for the next observable event in $(\Sigma_{o} - \Sigma_{s,E}) \cup \Sigma_{s,E}^{\#}$ or issue a control command.
    \item $q^{issue}$ is a state denoting that the supervisor has just issued a control command. At this state, the supervisor will wait for the next observable event in $(\Sigma_{o} - \Sigma_{s,E}) \cup \Sigma_{s,E}^{\#}$.
\end{itemize}
In the event set, any $\gamma \in \Gamma$ denotes the event of issuing a control command $\gamma$ by the supervisor.

For the (partial) transition function $\xi_{sc}$, 
\begin{itemize}
\setlength{\itemsep}{3pt}
\setlength{\parsep}{0pt}
\setlength{\parskip}{0pt}
    \item Cases 1 and 2 say that, at state $q_{sc}^{init}$, the supervisor would make a transition to state $q^{issue}$ only after it issues a control command $\gamma \in \Gamma$. If any other event $\sigma \in \Sigma_{sc} - \Gamma$ happens, the supervisor would only do a self-loop transition. 
    These two cases model the situation that the supervisor could either immediately issue a control command or wait for the next observation when 1) the supervisor has not issued any control command since the system initiates, or 2) the supervisor has observed at least one event $\sigma \in (\Sigma_{o} - \Sigma_{s,E}) \cup \Sigma_{s,E}^{\#}$ since it issues the last control command.
    \item Cases 3 and 4 say that, at state $q^{issue}$, since the supervisor has just issued a control command, it would not issue a control command again until receiving a new observation. Thus, at state $q^{issue}$, no $\gamma \in \Gamma$ is defined and the supervisor would make a transition to state $q_{sc}^{init}$ only after it observes an event $\sigma \in (\Sigma_{o} - \Sigma_{s,E}) \cup \Sigma_{s,E}^{\#}$. If any other event $\sigma \in \Sigma_{uo} \cup \Sigma_{s,E} \cup \{stop, decode\}$ happens, the supervisor would only do a self-loop transition since such events are unobservable to the supervisor.
\end{itemize}
Based on the model of $SC$, we have $|Q_{sc}| = 2$.

\textbf{Supervisor:} The supervisor is modeled as a finite state automaton $S$.
\[
S = (Q_{s}, \Sigma_{s}, \xi_{s}, q_{s}^{init}, Q_{s,m})
\]
where $\Sigma_{s} = \Sigma_{sc} = \Sigma \cup \Sigma_{s,E}^{\#} \cup \Gamma \cup \{stop, decode\}$, that satisfies the following constraints:
\begin{itemize}
\setlength{\itemsep}{3pt}
\setlength{\parsep}{0pt}
\setlength{\parskip}{0pt}
    \item (S-controllability) For any state $q \in Q_{s}$ and any $\sigma \in \Sigma_{s,uc}: = \Sigma_{s} - \Sigma_{s,c} = \Sigma_{s} - \Gamma$, $\xi_{s}(q, \sigma)!$.
    \item (S-observability) For any state $q \in Q_{s}$ and any $\sigma \in \Sigma_{s,uo}: = \Sigma_{s} - \Sigma_{s,o} = \Sigma_{s} - ((\Sigma_{o} - \Sigma_{s,E}) \cup \Sigma_{s,E}^{\#} \cup \Gamma)$, if $\xi_{s}(q, \sigma)!$, then $\xi_{s}(q, \sigma) = q$.
\end{itemize}
S-controllability states that the supervisor can only disable events in $\Sigma_{s,c} = \Gamma$. S-observability states that the supervisor can only make a state change after observing events in $\Sigma_{s,o} = (\Sigma_{o} - \Sigma_{s,E}) \cup \Sigma_{s,E}^{\#} \cup \Gamma$. In this work, by construction, all the controllable events for the supervisor are also observable to the supervisor. In the following text, we shall refer to $(\Sigma_{s,c}, \Sigma_{s,o})$ as the supervisor-control constraint.

\subsection{Command execution component}
\label{subsec:Command execution}
\begin{figure}[htbp]
\begin{center}
\includegraphics[height=2.5cm]{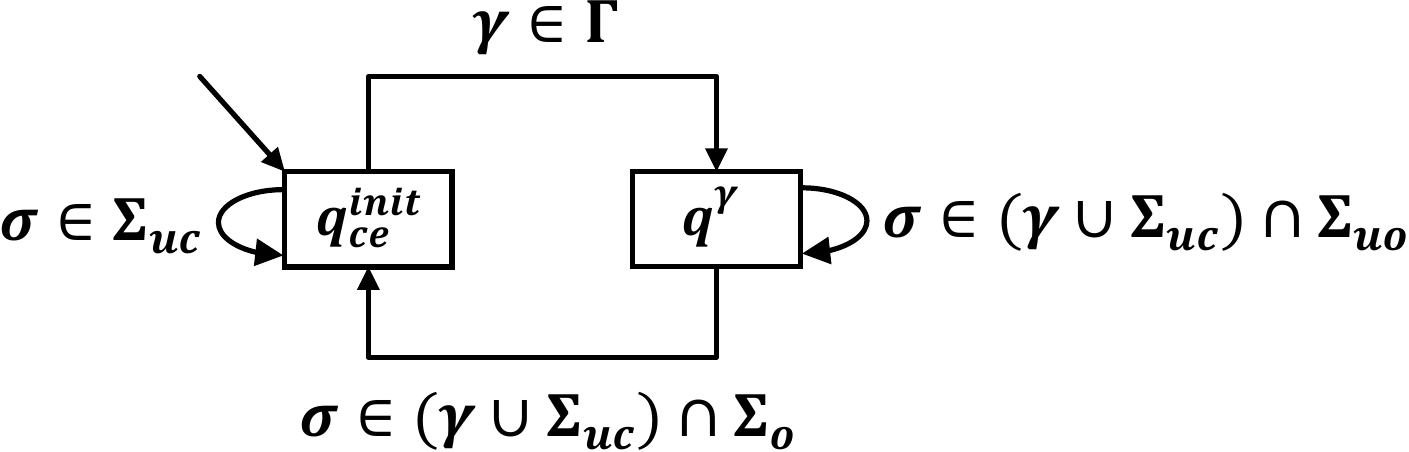}   
\caption{The (schematic) model for command execution}
\label{fig:The (schematic) model for command execution}
\end{center}        
\end{figure}

The command execution automaton serves to explicitly describe the execution phase of a control command, where the procedure from using a control command to executing an event is shown. Since the number of all possible control commands issued by the supervisor is finite\footnote{Since $\Gamma = 2^{\Sigma_{c}} - \{\varnothing\}$, the total number of control commands is upper bounded by $2^{|\Sigma_{c}|} - 1$, which is finite.}, the command execution component can be modeled as a finite state automaton $CE$, which is illustrated in Fig. \ref{fig:The (schematic) model for command execution}. 
\[
CE = (Q_{ce}, \Sigma_{ce}, \xi_{ce}, q_{ce}^{init}, Q_{ce,m})
\]
\begin{itemize}
\setlength{\itemsep}{3pt}
\setlength{\parsep}{0pt}
\setlength{\parskip}{0pt}
    \item $Q_{ce} = \{q^{\gamma}|\gamma \in \Gamma\} \cup \{q_{ce}^{init}\}$
    \item $\Sigma_{ce} = \Gamma \cup \Sigma$
    \item $\xi_{ce}: Q_{ce} \times \Sigma_{ce} \rightarrow Q_{ce}$
    \item $Q_{ce,m} = \{q_{ce}^{init}\}$
\end{itemize}
The (partial) transition function $\xi_{ce}$ is defined as follows:
\begin{enumerate}[1.]
\setlength{\itemsep}{3pt}
\setlength{\parsep}{0pt}
\setlength{\parskip}{0pt}
    \item For any $\gamma \in \Gamma$, $\xi_{ce}(q_{ce}^{init}, \gamma) = q^{\gamma}$.
    \item For any $\sigma \in (\gamma \cup \Sigma_{uc}) \cap \Sigma_{uo}$, $\xi_{ce}(q^{\gamma}, \sigma) = q^{\gamma}$.
    \item For any $\sigma \in (\gamma \cup \Sigma_{uc}) \cap \Sigma_{o}$, $\xi_{ce}(q^{\gamma}, \sigma) = q_{ce}^{init}$.
    \item For any $\sigma \in \Sigma_{uc}$, $\xi_{ce}(q_{ce}^{init}, \sigma) = q_{ce}^{init}$.
\end{enumerate}

Next, we shall present some explanations for the model $CE$. In the state set $Q_{ce}$, 
\begin{itemize}
\setlength{\itemsep}{3pt}
\setlength{\parsep}{0pt}
\setlength{\parskip}{0pt}
    \item $q_{ce}^{init}$ is a state denoting that the command execution component is not using any control command. At this state, the command execution component is waiting for the arrival of a control command issued by the supervisor. It is noteworthy that at state $q_{ce}^{init}$, any uncontrollable event is always allowed to be executed.
    \item $q^{\gamma} \in Q_{ce} (\gamma \in \Gamma)$ is a state denoting that the command execution component has just received the control command $\gamma$.
\end{itemize}
For the (partial) transition function $\xi_{ce}$,
\begin{itemize}
\setlength{\itemsep}{3pt}
\setlength{\parsep}{0pt}
\setlength{\parskip}{0pt}
    \item Case 1 says that, at state $q_{ce}^{init}$, if the command execution component receives a control command $\gamma$ issued by the supervisor, then it will transit to the state $q^{\gamma}$ and be ready to use $\gamma$.
    \item Case 2 says that, at state $q^{\gamma}$, if any event $\sigma \in (\gamma \cup \Sigma_{uc}) \cap \Sigma_{uo}$ is executed by the command execution component, then the command execution component will reuse the control command $\gamma$.
    \item Case 3 says that, at state $q^{\gamma}$, if any event $\sigma \in (\gamma \cup \Sigma_{uc}) \cap \Sigma_{o}$ is executed by the command execution component, then it will transit back to the state $q_{ce}^{init}$ and wait for the next control command to be issued by the supervisor.
    \item Case 4 says that, at state $q_{ce}^{init}$, any uncontrollable event $\sigma \in \Sigma_{uc}$ can be executed since uncontrollable events are always allowed to be fired.
\end{itemize}
Based on the model of $CE$, we have $|Q_{ce}| = 2^{|\Sigma_{c}|}$. 

\subsection{Plant}
\label{subsec:Plant}
Plant $G$ is modeled as a finite state automaton
\[
G = (Q, \Sigma, \xi_{G}, q^{init}, Q_{m})
\]
where the set of secret states in plant $G$ is denoted as $Q_{sec} \subseteq Q$, the set of bad states to avoid in the plant $G$ is denoted as $Q_{avoid} \subseteq Q$, the set of blocking states  in plant $G$ is denoted as $Q_{block} := \{q \in Q| (\forall s \in \Sigma^{*}) \, \xi(q, s)! \Rightarrow \xi(q, s) \notin Q_{m}\} \subseteq Q$.

In this work, we consider current-state opacity (CSO), which is defined as follows.

\emph{Definition III.1 (CSO)} \cite{Saboori2007SBO} $G = (Q, \Sigma, \xi_{G}, q^{init}, Q_{m})$ is CSO w.r.t. projection $P$ and the set of secret states $Q_{sec} \subseteq Q$ if 
\[
\begin{aligned}
& \forall t \in L_{S} := \{t \in L(G)|\xi_{G}(q^{init}, t) \in Q_{sec}\},  \exists t^{'} \in L_{NS} \\ & := \{t \in L(G)|\xi_{G}(q^{init}, t) \in (Q \backslash Q_{sec})\} , P(t) = P(t')
\end{aligned}
\]


\subsection{Intruder}
\label{subsec:Intruder}

As illustrated in Fig. \ref{fig:Supervisory control architecture under edit function}, the intruder is an external observer that aims to infer the system secret based on its observations. In this work, the assumptions about the intruder are given as follows:
\begin{itemize}
\setlength{\itemsep}{3pt}
\setlength{\parsep}{0pt}
\setlength{\parskip}{0pt}
    \item The set of observable events for the intruder is denoted as $\Sigma_{o,I} \subseteq \Sigma$. It is noteworthy that $\Sigma_{o,I}$ might be different from $\Sigma_{o}$, the set of observable events for the supervisor, and $\Sigma_{o,E}$, the set of observable events for the edit function. 
    \item The intruder only has the full knowledge of the structure of the plant $G$ and does not know the model of the supervisor and the specification.
\end{itemize}
Based on the above assumptions, it is known that,
\begin{itemize}
\setlength{\itemsep}{3pt}
\setlength{\parsep}{0pt}
\setlength{\parskip}{0pt}
    \item Due to the existence of the edit function, all the events in $\Sigma_{s,E}$ have been relabelled as the copies in $\Sigma_{s,E}^{\#}$. Thus, in the modeling, the intruder could only observe events in $(\Sigma_{o,I} - \Sigma_{s,E}) \cup (\Sigma_{o,I} \cap \Sigma_{s,E})^{\#}$. 
    \item Since the structure of the plant $G$ is a prior knowledge of the intruder, the intruder is able to compare its online observation sequences during the system running with the ones that could have been observed under the absence of an edit function. Once the information inconsistency happens, the intruder will conclude the existence of the edit function. 
\end{itemize}
Thus, under the supervision of $S$  and in the presence of $E$, the following goals should be achieved:
\begin{enumerate}[1.]
\setlength{\itemsep}{3pt}
\setlength{\parsep}{0pt}
\setlength{\parskip}{0pt}
    \item Plant $G$ would never reach the state in $Q_{avoid}$ and the closed-loop system behavior is nonblocking.
    \item The intruder would never infer that plant $G$ has reached a secret state in $Q_{sec}$. 
    \item The existence of the edit function is never exposed to the intruder. 
\end{enumerate}

Next, we shall explain how to model the intruder, which consists of the following two steps.

\vspace{0.1cm}
\textbf{Step 1:} 
On one hand, the intruder could only observe events in $\Sigma_{o,I}$. On the other hand, the intruder could discover the existence of an edit function based on its online observations.
To capture the above-mentioned two features of the intruder, we construct the following finite state automaton 
\[
P_{\Sigma_{o,I}}(G) = (Q_{temp}, \Sigma_{temp}, \xi_{temp}, q_{temp}^{init})
\]
\begin{itemize}
\setlength{\itemsep}{3pt}
\setlength{\parsep}{0pt}
\setlength{\parskip}{0pt}
    \item $Q_{temp} = 2^{Q}$
    \item $\Sigma_{temp} = \Sigma$
    \item $\xi_{temp}: Q_{temp} \times \Sigma_{temp} \rightarrow Q_{temp}$
    \item $q_{temp}^{init} = UR_{G, \Sigma_{temp} - \Sigma_{o,I}}(q^{init})$
\end{itemize}
 $P_{\Sigma_{o,I}}(G)$ is essentially a state estimator, where we have the following two facts: 1) once $P_{\Sigma_{o,I}}(G)$ transits to a state in $2^{Q_{sec}} - \{\varnothing\}$, the intruder infers the secret state of the plant $G$; 2) once $P_{\Sigma_{o,I}}(G)$ transits to state $\varnothing$, the existence of the edit function is discovered by the intruder.

\vspace{0.2cm}
\textbf{Step 2:} We shall make some minor modifications on $P_{\Sigma_{o,I}}(G)$ due to the following reasons: 1) The intruder can infer the secret state of plant $G$ after  $P_{\Sigma_{o,I}}(G)$ transits to a state in $2^{Q_{sec}} - \{\varnothing\}$. We model it in such a way that once the intruder infers that the plant $G$ has reached a secret state, it would transit to a new state $q^{unsafe}$ by a new transition labelled with the event $decode$, which will be introduced below;
2) Since the intruder could only observe events in $(\Sigma_{o,I} - \Sigma_{s,E}) \cup (\Sigma_{o,I} \cap \Sigma_{s,E})^{\#}$, we need to replace any $\sigma \in \Sigma_{s,E}$ by $\sigma^{\#}$ in the model of the intruder. Then, based on $P_{\Sigma_{o,I}}(G)$, the model of the intruder $I$ is generated by the following procedure.
\[
I = (Q_{i}, \Sigma_{i}, \xi_{i}, q_{i}^{init}, Q_{i,m})
\]
\begin{enumerate}[1.]
\setlength{\itemsep}{3pt}
\setlength{\parsep}{0pt}
\setlength{\parskip}{0pt}
    \item $Q_{i} = Q_{temp} \cup \{q^{unsafe}\}$
    \item $\Sigma_{i} = (\Sigma_{temp} - \Sigma_{s,E}) \cup \Sigma_{s,E}^{\#} \cup \{decode\} = (\Sigma - \Sigma_{s,E}) \cup \Sigma_{s,E}^{\#} \cup \{decode\}$
    \item  $(\forall q \in Q_{i}) \, q \in 2^{Q_{sec}} - \{\varnothing\} \Leftrightarrow \xi_{i}(q, decode) = q^{unsafe}$
    \item $(\forall q, q' \in Q_{i})(\forall \sigma \in \Sigma_{s,E}) \, \xi_{temp}(q, \sigma) = q' \Leftrightarrow \xi_{i}(q, \sigma^{\#}) = q'$
    \item $(\forall q, q' \in Q_{i})(\forall \sigma \in \Sigma - \Sigma_{s,E}) \, \xi_{temp}(q, \sigma) = q' \Leftrightarrow \xi_{i}(q, \sigma) = q'$
    \item $(\forall q \in \{\varnothing, q^{unsafe}\})(\forall \sigma \in (\Sigma - \Sigma_{s,E}) \cup \Sigma_{s,E}^{\#})\, \xi_{i}(q,\sigma)\\ = q$
    \item $q_{i}^{init} = q_{temp}^{init}$
    \item $Q_{i,m} = Q_{temp} - \{\varnothing\}$
\end{enumerate}
We shall give some explanations for the above procedure. 
\begin{itemize}
\setlength{\itemsep}{3pt}
\setlength{\parsep}{0pt}
\setlength{\parskip}{0pt}
    \item In Step 1 and Step 2, a new state $q^{unsafe}$ and a new event $decode$, denoting that the intruder infers that plant $G$ has reached a secret state, is added to the state set and the event set, respectively, which generates the new state set $Q_{i}$ and new event set $\Sigma_{i}$, respectively.
    \item In Step 3, to encode the situation that the intruder infers that plant $G$ has reached a secret state at each state $q \in 2^{Q_{sec}} - \{\varnothing\}$, a new outgoing transition is added to state $q$, such that $q \in 2^{Q_{sec}} - \{\varnothing\} \Leftrightarrow \xi_{i}(q, decode) = q^{unsafe}$. In this case, as long as the intruder transits to state $q^{unsafe}$, the secret state of plant $G$ has been inferred. Since the event $decode$ is uncontrollable to the edit function and the supervisor, thus, to enforce current-state opacity, the intruder should never transit to any state $q \in 2^{Q_{sec}} - \{\varnothing\}$ under the supervision of $S$ in the presence of $E$.
    \item In Steps 4 and 5, all the transitions labelled by events in $\Sigma_{s,E}$ are replaced with the relabelled copies in $\Sigma_{s,E}^{\#}$ while other transitions remain the same.
    \item In Step 6, any event in $(\Sigma - \Sigma_{s,E}) \cup \Sigma_{s,E}^{\#}$ is defined as a self-loop at state $\varnothing$ or $q^{unsafe}$ since now any event execution or any further observation at the intruder would not change the fact that it has already either inferred the system secret or discovered the existence of the edit function.
\end{itemize}

Based on the constructed model of the intruder, from the point view of the edit function and supervisor, it should avoid the transitions to state $\varnothing$ and $q^{unsafe}$ in $I$. Based on the model of $I$, we have $|Q_{i}| \leq 2^{|Q|}+1$.

\section{Co-Synthesis of Edit Function and Supervisor for Opacity Enforcement}
\label{sec:Co-Synthesis of Edit Function and Supervisor for Opacity Enforcement}

In this section, firstly, based on the component models presented in Section \ref{sec:Component Models of DES under Edit Function and Supervisor}, we shall formalize the closed-loop behavior of the system under edit function, supervisor, and intruder. Based on the closed-loop behavior, we shall introduce several definitions, including opacity and covertness. Then, we shall solve the co-synthesis problem of edit function and supervisor for opacity enforcement by modeling it as a distributed supervisor synthesis problem in the  Ramadge-Wonham supervisory control framework.

\subsection{Solution Methodology}
\label{subsec:Solution Methodology}

In Fig. \ref{fig:Supervisory control architecture under edit function}, given the plant $G$, the command execution component $CE$, the edit constraints $EC$, the supervisor constraints $SC$, the intruder $I$, the edit function $E$, and the supervisor $S$, the closed-loop  system, defined as $\mathcal{B}$, is the synchronous product given as follows:
\[
\begin{aligned}
\mathcal{B} & = G||CE||EC||SC||I||E||S \\ & = (Q_{b}, \Sigma_{b}, \xi_{b}, q_{b}^{init}, Q_{b,m})
\end{aligned}
\]
\begin{itemize}
\setlength{\itemsep}{3pt}
\setlength{\parsep}{0pt}
\setlength{\parskip}{0pt}
    \item $Q_{b} = Q \times Q_{ce} \times Q_{ec} \times Q_{sc} \times Q_{i} \times Q_{e} \times Q_{s}$
    \item $\Sigma_{b} = \Sigma \cup \Sigma_{ce} \cup \Sigma_{ec} \cup \Sigma_{sc} \cup \Sigma_{i} \cup \Sigma_{e} \cup \Sigma_{s} = \Sigma \cup \Sigma_{s,E}^{\#} \cup \Gamma \cup \{stop, decode\}$ 
    \item $\xi_{b}: Q_{b} \times \Sigma_{b} \rightarrow Q_{b}$
    \item $q_{b}^{init} = (q^{init}, q_{ce}^{init}, q_{ec}^{init}, q_{sc}^{init}, q_{i}^{init}, q_{e}^{init}, q_{s}^{init})$
    \item $Q_{b,m} = Q_{m} \times Q_{ce,m} \times Q_{ec,m} \times Q_{sc} \times Q_{i,m} \times Q_{e,m} \times Q_{s,m}$ 
\end{itemize}

Next, based on the closed-loop system $\mathcal{B}$, we shall present several definitions regarding the properties of edit function and supervisor for opacity enforcement. In the following text, for convenience, we shall refer to the edit function combined with the supervisor as an edit function-supervisor pair.

\emph{Definition IV.1. (Opacity):} Given any plant $G$, command execution component  $CE$, edit constraints $EC$, supervisor constraints $SC$, and intruder $I$, the edit function $E$ combined with the supervisor $S$ is an opaque edit function-supervisor pair w.r.t. the edit function-control constraint $(\Sigma_{e,c}, \Sigma_{e,o})$ and supervisor-control constraint $(\Sigma_{s,c}, \Sigma_{s,o})$ for $G$, $CE$, $EC$, $SC$, and $I$ if any state in 
\[
Q_{unsafe} = \{(q, q_{ce}, q_{ec}, q_{sc}, q_{i}, q_{e}, q_{s}) \in Q_{b}|\, q_{i} = q^{unsafe}\}
\]
is not reachable in $\mathcal{B} = G||CE||EC||SC||I||E||S$.

\emph{Definition IV.2. (Covertness):} Given any plant $G$, command execution component $CE$, edit constraints $EC$, supervisor constraints $SC$, and intruder $I$, the edit function $E$ combined with the supervisor $S$ is a covert edit function-supervisor pair w.r.t. the edit function-control constraint $(\Sigma_{e,c}, \Sigma_{e,o})$ and supervisor-control constraint $(\Sigma_{s,c}, \Sigma_{s,o})$ for $G$, $CE$, $EC$, $SC$, and $I$ if any state in 
\[
Q_{bad} = \{(q, q_{ce}, q_{ec}, q_{sc}, q_{i}, q_{e}, q_{s}) \in Q_{b}|\, q_{i} = \varnothing\}
\]
is not reachable in $\mathcal{B} = G||CE||EC||SC||I||E||S$.

Next, we shall further explain our approach in modeling the problem of co-synthesis of edit function and supervisor for opacity enforcement as a distributed Ramadge-Wonham supervisory control problem.
Since the closed-loop  system is $\mathcal{B} = G||CE||EC||SC||I||E||S$, we can view
\[
\mathcal{P} = G||CE||EC||SC||I = (Q_{\mathcal{P}}, \Sigma_{\mathcal{P}}, \xi_{\mathcal{P}}, q_{\mathcal{P}}^{init}, Q_{\mathcal{P},m})
\]
as the new plant and treat $E$ and $S$ as the distributed supervisor to be synthesized over the distributed control architecture $\mathcal{A}=((\Sigma_{e, c}, \Sigma_{e, o}), (\Sigma_{s, c}, \Sigma_{s, o}))$. Our goal is to synthesize $E$ and $S$ such that
\begin{itemize}
\setlength{\itemsep}{3pt}
\setlength{\parsep}{0pt}
\setlength{\parskip}{0pt}
    \item $\mathcal{B}$ is nonblocking and plant $G$ would never reach any state in $Q_{avoid}$.
    \item $E$ combined with $S$ is an opaque edit function-supervisor pair w.r.t. the edit function-control constraint $(\Sigma_{e,c}, \Sigma_{e,o})$ and supervisor-control constraint $(\Sigma_{s,c}, \Sigma_{s,o})$ for $G$, $CE$, $EC$, $SC$, and $I$.
    \item $E$ combined with $S$ is a covert edit function-supervisor pair w.r.t. the edit function-control constraint $(\Sigma_{e,c}, \Sigma_{e,o})$ and supervisor-control constraint $(\Sigma_{s,c}, \Sigma_{s,o})$ for $G$, $CE$, $EC$, $SC$, and $I$.
\end{itemize}

Before we present our heuristic for solving the co-synthesis problem, we briefly discussed about some previous approaches for addressing the distributed supervisor synthesis problem, which is known to be undecidable in general \cite{Lin2016Distributed}-\cite{Thistle2005Distributed}: 1) \cite{Komenda2014CoordinationControl} proposes a distributed synthesis approach by adopting an coordinator, which receives part of the partial observations of the subsystems and serves to satisfy the global specification and nonblockingness. Nevertheless, in the architecture shown in Fig. \ref{fig:Supervisory control architecture under edit function}, we do not allow such a coordinator for the privacy-preserving control problem. Thus, the approach in \cite{Komenda2014CoordinationControl} is not applicable for the synthesis problem to be solved in our work. 
2) \cite{wonham2015supervisory} summarizes the supervisor localization algorithm for the distributed control for DES. However, this algorithm often needs to lift the observation alphabets of the local supervisors,  which is not suitable for the distributed synthesis problem to be solved in our work, which has a fixed distributed control architecture. 
3) \cite{Su2010Aggregative} proposes an aggregative synthesis approach that computes nonblocking distributed supervisors, which always tries to synthesize a nonblocking supervisor at each step. However, for the distributed supervisor synthesis problem to be solved in this work, since the events denoting the edit operations (respectively, the sending of control commands) are uncontrollable to the supervisor (respectively, edit function), no matter whether we synthesize $S$ or $E$ first, the algorithm in \cite{Su2010Aggregative} is very likely to output an empty solution at the first step.

In this work, we take the special structure of this distributed supervisor synthesis problem into consideration and propose two heuristics to generate the desired $E$ and $S$ to achieve the safety, opacity, covertness, and nonblockingness goal, where one heuristic first synthesizes $S$ and then synthesizes $E$, and the other heuristic first synthesizes $E$ and then synthesizes $S$. The details of these two heuristics would be explained in Section \ref{subsec:Synthesis_S_E} and Section \ref{subsec:Synthesis_E_S}, respectively.



 \subsection{Incremental synthesis: first $S$ and then $E$}
\label{subsec:Synthesis_S_E}
In this heuristic, we first synthesize the supervisor $S$ to ensure the safety of $G$ and the marker-reachability of the closed-loop system, and then we synthesize the edit function $E$ to ensure the opacity, covertness and nonblockingness.  The details of the synthesis procedure are as follows: 

\vspace{0.1cm}

\noindent \textbf{Procedure 1:}
\begin{enumerate}[1.]
\setlength{\itemsep}{3pt}
\setlength{\parsep}{0pt}
\setlength{\parskip}{0pt}
    \item Compute $\mathcal{P} = G||CE||EC||SC||I = (Q_{\mathcal{P}}, \Sigma_{\mathcal{P}}, \xi_{\mathcal{P}}, q_{\mathcal{P}}^{init},\\ Q_{\mathcal{P},m})$. 
    \item Generate $\mathcal{P}_{S}^{0} := (Q_{\mathcal{P}_{S}^{0}}, \Sigma_{\mathcal{P}_{S}^{0}}, \xi_{\mathcal{P}_{S}^{0}}, q_{\mathcal{P}_{S}^{0}}^{init}, Q_{\mathcal{P}_{S}^{0},m})$
    \begin{itemize}
    \setlength{\itemsep}{3pt}
    \setlength{\parsep}{0pt}
    \setlength{\parskip}{0pt}
        \item $Q_{\mathcal{P}_{S}^{0}} := Q_{\mathcal{P}} - Q_{1} - Q_{2} - Q_{3}$
        \begin{itemize}
        \setlength{\itemsep}{3pt}
        \setlength{\parsep}{0pt}
        \setlength{\parskip}{0pt}
            \item $Q_{1} := \{(q, q_{ce}, q_{ec}, q_{sc}, q_{i}) \in Q_{\mathcal{P}}|\, q \in Q_{avoid}\}$
            \item $Q_{2} := \{(q, q_{ce}, q_{ec}, q_{sc}, q_{i}) \in Q_{\mathcal{P}}|\, q \in Q_{block}\}$
            \item $Q_{3} := \{(q, q_{ce}, q_{ec}, q_{sc}, q_{i}) \in Q_{\mathcal{P}}|\, q \notin Q_{m} \wedge q_{ce} = q^{\gamma} \neq q_{ce}^{init} \wedge (\gamma \cup \Sigma_{uc}) \cap En_{G}(q) = \varnothing\}$ 
        \end{itemize}
        \item $\Sigma_{\mathcal{P}_{S}^{0}} := \Sigma_{\mathcal{P}}$
        \item $(\forall q, q^{'} \in Q_{\mathcal{P}_{S}^{0}})(\forall \sigma \in \Sigma_{\mathcal{P}_{S}^{0}}) \, \xi_{\mathcal{P}}(q, \sigma) = q^{'} \Leftrightarrow \xi_{\mathcal{P}_{S}^{0}}(q, \sigma) = q^{'}$
        \item $q_{\mathcal{P}_{S}^{0}}^{init} := q_{\mathcal{P}}^{init}$
        \item $Q_{\mathcal{P}_{S}^{0},m} := Q_{\mathcal{P},m} - Q_{1} - Q_{2} - Q_{3}$
    \end{itemize}
    \item Synthesize a supervisor $S_{0} = (Q_{s,0}, \Sigma_{s,0}, \xi_{s,0}, q_{s,0}^{init}, \\Q_{s,0,m})$ over the supervisor-control constraint $(\Sigma_{s,c}, \Sigma_{s,o})$ by treating $\mathcal{P}$ as the plant and $\mathcal{P}_{S}^{0}$ as the requirement such that $\mathcal{P}||S_0$ is marker-reachable and safe w.r.t. $\mathcal{P}_{S}^{0}$. If $S_{0}$ exists, go to Step 4; otherwise, end the procedure.
    \item Let $k \leftarrow 0$.
    \item Compute $\mathcal{P}||S_{k} = (Q_{\mathcal{P}||S_{k}}, \Sigma_{\mathcal{P}||S_{k}}, \xi_{\mathcal{P}||S_{k}}, q_{\mathcal{P}||S_{k}}^{init},\\ Q_{\mathcal{P}||S_{k},m})$.
    \item Generate $Q_{del} := \{q \in Q_{\mathcal{P}_{S}^{k}}|\, C_{1} \vee C_{2}\}$, where
    \[
    \begin{aligned}
    &C_{1} := \begin{aligned}
    &(\exists (q,q_{s}) \in Q_{\mathcal{P}||S_{k}})(\forall t \in \Sigma_{\mathcal{P}||S_{k}}^{*})\,  \xi_{\mathcal{P}||S_{k}}((q,q_{s}),t) \\ & \notin Q_{m} \times Q_{ce} \times Q_{ec} \times Q_{sc} \times Q_{i} \times Q_{s,k} 
    \end{aligned}
    \\ &C_{2} := (\forall q_{s} \in Q_{s,k})\, (q, q_{s}) \notin Q_{\mathcal{P}||S_{k}}
    \end{aligned}
    \]
    If $Q_{del} \neq \varnothing$, go to Step 7; otherwise, go to Step 9 and treat $S_{k}$ as the desired supervisor $S$, denoted as $S := S_{k} = (Q_{s}, \Sigma_{s}, \xi_{s}, q_{s}^{init}, Q_{s,m})$.
    \item Generate $\mathcal{P}_{S}^{k+1} := (Q_{\mathcal{P}_{S}^{k+1}}, \Sigma_{\mathcal{P}_{S}^{k+1}}, \xi_{\mathcal{P}_{S}^{k+1}}, q_{\mathcal{P}_{S}^{k+1}}^{init}, Q_{\mathcal{P}_{S}^{k+1},m})$
    \begin{itemize}
    \setlength{\itemsep}{3pt}
    \setlength{\parsep}{0pt}
    \setlength{\parskip}{0pt}
        \item $Q_{\mathcal{P}_{S}^{k+1}} := Q_{\mathcal{P}_{S}^{k}} - Q_{del}$
        \item $\Sigma_{\mathcal{P}_{S}^{k+1}} := \Sigma_{\mathcal{P}_{S}^{k}}$
        \item $(\forall q, q^{'} \in Q_{\mathcal{P}_{S}^{k+1}})(\forall \sigma \in \Sigma_{\mathcal{P}_{S}^{k+1}}) \, \xi_{\mathcal{P}_{S}^{k}}(q, \sigma) = q^{'} \Leftrightarrow \xi_{\mathcal{P}_{S}^{k+1}}(q, \sigma) = q^{'}$
        \item $q_{\mathcal{P}_{S}^{k+1}}^{init} := q_{\mathcal{P}_{S}^{k}}^{init}$
        \item $Q_{\mathcal{P}_{S}^{k+1},m} := Q_{\mathcal{P}_{S}^{k},m} - Q_{del}$
    \end{itemize}
    \item Synthesize a supervisor $S_{k+1} = (Q_{s,k+1}, \Sigma_{s,k+1}, \xi_{s,k+1},\\ q_{s,k+1}^{init}, Q_{s,k+1,m})$ over the supervisor-control constraint $(\Sigma_{s,c}, \Sigma_{s,o})$ by treating $\mathcal{P}$ as the plant and $\mathcal{P}_{S}^{k+1}$ as the requirement such that $\mathcal{P}||S_{k+1}$ is marker-reachable and safe w.r.t. $\mathcal{P}_{S}^{k+1}$. If $S_{k+1}$ exists, let $k \leftarrow k+1$ and go to Step 5; otherwise, end the procedure.  
    \item Compute $\mathcal{P}_{E} = G||CE||EC||SC||I||S = (Q_{\mathcal{P}_{E}}, \Sigma_{\mathcal{P}_{E}}, \\ \xi_{\mathcal{P}_{E}}, q_{\mathcal{P}_{E}}^{init}, Q_{\mathcal{P}_{E},m})$
    \item Generate $\mathcal{P}_{E}^{r} := (Q_{\mathcal{P}_{E}^{r}}, \Sigma_{\mathcal{P}_{E}^{r}}, \xi_{\mathcal{P}_{E}^{r}}, q_{\mathcal{P}_{E}^{r}}^{init}, Q_{\mathcal{P}_{E}^{r},m})$
    \begin{itemize}
    \setlength{\itemsep}{3pt}
    \setlength{\parsep}{0pt}
    \setlength{\parskip}{0pt}
        \item $Q_{\mathcal{P}_{E}^{r}} := Q_{\mathcal{P}_{E}} - Q_{4}$
        \begin{itemize}
        \setlength{\itemsep}{3pt}
        \setlength{\parsep}{0pt}
        \setlength{\parskip}{0pt}
            \item $Q_{4} := \{(q, q_{ce}, q_{ec}, q_{sc}, q_{i}, q_{s}) \in Q_{\mathcal{P}_{E}}|\, q_{i} = q^{unsafe} \vee q_{i} = \varnothing\}$
        \end{itemize}
        \item $\Sigma_{\mathcal{P}_{E}^{r}} := \Sigma_{\mathcal{P}_{E}}$
        \item $(\forall q, q' \in Q_{\mathcal{P}_{E}^{r}})(\forall \sigma \in \Sigma_{\mathcal{P}_{E}^{r}}) \, \xi_{\mathcal{P}_{E}}(q, \sigma) = q' \Leftrightarrow \xi_{\mathcal{P}_{E}^{r}}(q, \sigma) = q'$
        \item $q_{\mathcal{P}_{E}^{r}}^{init} := q_{\mathcal{P}_{E}}^{init}$
        \item $Q_{\mathcal{P}_{E}^{r},m} := Q_{\mathcal{P}_{E},m} - Q_{4}$
    \end{itemize}
    \item Synthesize a supervisor $E = (Q_{e}, \Sigma_{e}, \xi_{e}, q_{e}^{init}, Q_{e,m})$ over the edit function-control constraint $(\Sigma_{e,c}, \Sigma_{e,o})$ by treating $\mathcal{P}_{E}$ as the plant and $\mathcal{P}_{E}^{r}$ as the requirement such that $\mathcal{P}_{E}||E$ is nonblocking and safe w.r.t. $\mathcal{P}_{E}^{r}$.
\end{enumerate}

In the above procedure, Steps 1-8 are dedicated to synthesizing the supervisor and Steps 9-11 are dedicated to synthesizing the edit function based on the synthesized supervisor. In the part regarding the synthesis of the supervisor, for Steps 1-3, $\mathcal{P}$ and $\mathcal{P}_{S}^{0}$ are treated as the plant and the requirement, respectively. The requirement $\mathcal{P}_{S}^{0}$ is generated by removing three kinds of states in $\mathcal{P}$: 1) the states where the plant $G$ reaches a state in $Q_{avoid}$, denoted by $Q_{1}$ in Step 2; 2) the states where the plant $G$ reaches a state in $Q_{block}$, denoted by $Q_{2}$ in Step 2; 3) the states where the plant $G$ has not reached a marker state in $Q_{m}$ meanwhile the command execution component is using a control command $\gamma$ such that $\gamma \cup \Sigma_{uc}$ has no intersection with the enabled events of current state of $G$, denoted by $Q_{3}$ in Step 2. We need to delete such states because: 1) the first kind of states are the ``bad'' states that are not allowed by the user requirement and they should be avoided; 2) the second kind of states are those states where the  nonblockingness goal of $G$ already cannot be satisfied; 3) the third kind of states are those deadlocked, non-marked states where the supervisor issues some control commands that cannot be used by the plant. 
 Based on $\mathcal{P}_{S}^{0}$, at Step 3, we compute the supervisor $S_{0}$ that could ensure the safety w.r.t. $\mathcal{P}_{S}^{0}$ and the marker-reachability. At this step, the nonblockingness of the closed-loop system is hard to ensure since the events denoting edit operations are uncontrollable to the supervisor and can easily cause blockingness.

It is noteworthy that although $S_{0}$ could ensure the reachability of some marker states in the closed-loop system behavior, it is still possible that the blockingness\footnote{The blockingness here is in terms of the behavior of the plant $G$, not the closed-loop system behavior.} can happen in the plant $G$ under the supervision of $S_{0}$. If so, then the nonblockingness of the closed-loop system behavior can be hard to ensure when we synthesize the edit function based on $S_{0}$, since the events denoting the sending of control commands by the supervisor are uncontrollable to the edit function. Thus, to improve the possibility of finding a non-empty edit function, we need to iteratively perform the synthesis until the blockingness would not happen in the plant $G$ under the supervision of such a supervisor. We shall refer to Step 1-3 as the 0-th iteration. The iterative computations are given in Steps 5-8: Firstly, for the $(k+1)$-th iteration, at Step 5, we compute the synchronous product of $\mathcal{P}$ and $S_{k}$ synthesized at the $k$-th iteration. Then, at Step 6, we identify the state $q \in Q_{\mathcal{P}_{S}^{k}}$ that satisfy one of the following conditions:
\begin{itemize}
\setlength{\itemsep}{3pt}
\setlength{\parsep}{0pt}
\setlength{\parskip}{0pt}
    \item There exists a state $(q,q_{s})$ in $\mathcal{P}||S_{k}$ that cannot reach any state in $Q_{m} \times Q_{ce} \times Q_{ec} \times Q_{sc} \times Q_{i} \times Q_{s,k}$.
    \item For any state in $\mathcal{P}||S_{k}$, the tuple consisting of the first five terms of this state is not equal to $q$, denoted by $(q,q_{s}) \notin Q_{\mathcal{P}||S_{k}}$. 
\end{itemize}
The first condition corresponds to the situation that blockingness happens in terms of the behavior of plant $G$ under the supervision of $S_{k}$. The second condition corresponds to the situation that $\mathcal{P}$ would not transit to state $q$ under the supervision of $S_{k}$, thus, such state $q$ should also be avoided in the requirement at the $(k+1)$-th iteration. Any state in $Q_{\mathcal{P}_{S}^{k}}$ satisfying the above two conditions would be contained in $Q_{del}$.
If $Q_{del} \neq \varnothing$, then we need to remove such states in the requirement $\mathcal{P}_{S}^{k}$ to generate a new requirement $\mathcal{P}_{S}^{k+1}$ at Step 7, based on which we compute $S_{k+1}$ to ensure safety and reachability at Step 8. If $Q_{del} = \varnothing$, then $S_{k}$ is the desired supervisor and the procedure moves to Step 9.

In the part regarding the synthesis of the edit function (Steps 9-11), $\mathcal{P}_{E}$ and $\mathcal{P}_{E}^{r}$ are treated as the plant and requirement, respectively. The plant $\mathcal{P}_{E}$ is generated based on the synthesized supervisor $S$ in Steps 1-8. The requirement $\mathcal{P}_{E}^{r}$ is generated from $\mathcal{P}_{E}$ by removing the states where the intruder reaches the state $q^{unsafe}$ or $\varnothing$, implying that either the intruder has inferred the system secret or the existence of the edit function has been discovered, both of which should be avoided by the edit function. Finally, we compute the edit function $E$ that could satisfy the opacity, covertness, and nonblockingness at Step 11.

\emph{Theorem IV.1:} Given any plant $G$, command execution component $CE$, edit constraints $EC$, supervisor constraints $SC$, and intruder $I$, \textbf{Procedure 1} terminates within finite steps.

\emph{Proof:} To show this, we only need to check whether the iterative computation in Steps 5-8 can terminate within finite steps. Since the continuation of the iteration at Step 6 depends on whether $Q_{del}$ is equal to $\varnothing$, the worst case is that only one state is removed from $\mathcal{P}_{S}^{k}$ at each iteration. In addition,  $\mathcal{P}_{S}^{0}$ is a finite state automaton, which implies that the iterative computation in Steps 5-8 can always terminate within finite steps. This completes the proof. \hfill $\blacksquare$

\vspace{0.1cm}

\emph{Theorem IV.2:} Given any plant $G$, command execution component $CE$, edit constraints $EC$, supervisor constraints $SC$, and intruder $I$, the computed $S$ and $E$ in \textbf{Procedure 1}, if not empty, satisfy the following properties:
\begin{itemize}
\setlength{\itemsep}{3pt}
\setlength{\parsep}{0pt}
\setlength{\parskip}{0pt}
    \item $G||CE||EC||SC||I||E||S$ is nonblocking and any state in $\{(q, q_{ce}, q_{ec}, q_{sc}, q_{i}, q_{e}, q_{s}) \in Q \times Q_{ce} \times Q_{ec} \times Q_{sc} \times Q_{i} \times Q_{e} \times Q_{s}|\, q \in Q_{avoid}\}$ is not reachable in $G||CE||EC||SC||I||E||S$.
    \item $E$ combined with $S$ is an opaque edit function-supervisor pair w.r.t. the edit function-control constraint $(\Sigma_{e,c}, \Sigma_{e,o})$ and supervisor-control constraint $(\Sigma_{s,c}, \Sigma_{s,o})$ for $G$, $CE$, $EC$, $SC$, and $I$.
    \item $E$ combined with $S$ is a covert edit function-supervisor pair w.r.t. the edit function-control constraint $(\Sigma_{e,c}, \Sigma_{e,o})$ and supervisor-control constraint $(\Sigma_{s,c}, \Sigma_{s,o})$ for $G$, $CE$, $EC$, $SC$, and $I$.
\end{itemize}

\emph{Proof:} Based on Step 11 of \textbf{Procedure 1}, the synthesized $E$ should satisfy that $\mathcal{P}_{E}||E$ is nonblocking, that is, $G||CE||EC||SC||I||E||S$ is nonblocking. 
Based on Step 2 in \textbf{Procedure 1}, the set of states $Q_{1}$ has been removed in the requirement $\mathcal{P}_{S}^{0}$, i.e., they are treated as ``bad'' states in the synthesis of $S_{0}$. Thus, any state in $\{(q, q_{ce}, q_{ec}, q_{sc}, q_{i}, q_{e}, q_{s}) \in Q \times Q_{ce} \times Q_{ec} \times Q_{sc} \times Q_{i} \times Q_{e} \times Q_{s}|\, q \in Q_{avoid}\}$ is not reachable in $G||CE||EC||SC||I||E||S$. In addition, the set of states $Q_{4} := \{(q, q_{ce}, q_{ec}, q_{sc}, q_{i}, q_{s}) \in Q_{\mathcal{P}_{E}}|\, q_{i} = q^{unsafe} \vee q_{i} = \varnothing\}$ has been removed in the requirement $\mathcal{P}_{E}^{r}$, i.e., they are treated as ``bad'' states in the synthesis of $E$. Thus, any state in $\{(q, q_{ce}, q_{ec}, q_{sc}, q_{i}, q_{e}, q_{s}) \in Q \times Q_{ce} \times Q_{ec} \times Q_{sc} \times Q_{i} \times Q_{e} \times Q_{s}|\, q_{i} = q^{unsafe} \vee q_{i} = \varnothing\}$ is not reachable in the closed-loop system $G||CE||EC||SC||I||E||S$. Based on the definition IV.1 and IV.2, $E$ combined with $S$ is an opaque and covert edit function-supervisor pair w.r.t. the edit function-control constraint $(\Sigma_{e,c}, \Sigma_{e,o})$ and supervisor-control constraint $(\Sigma_{s,c}, \Sigma_{s,o})$ for $G$, $CE$, $EC$, $SC$, and $I$. This completes the proof. \hfill $\blacksquare$

Next, we shall analyze the computational complexity of \textbf{Procedure 1}.
In Steps 5-8, the worst case is that only one state is removed from the requirement at each iteration. Thus, by adopting the normality based synthesis approach in \cite{WangStateControl2018}, the complexity is 
\[
\begin{aligned}
&O(|\Sigma_{\mathcal{P}}|2^{|Q_{\mathcal{P}}|} + \dots + |\Sigma_{\mathcal{P}}| \times 2 + |\Sigma_{\mathcal{P}}||Q_{\mathcal{P}_{E}}|^{2}4^{|Q_{\mathcal{P}_{E}}|}) \\ = \, &O(2|\Sigma_{\mathcal{P}}|(2^{|Q_{\mathcal{P}}|}-1) + |\Sigma_{\mathcal{P}}||Q_{\mathcal{P}_{E}}|^{2}4^{|Q_{\mathcal{P}_{E}}|})\\
= \, &O(|\Sigma_{\mathcal{P}}||Q_{\mathcal{P}_{E}}|^{2}4^{|Q_{\mathcal{P}_{E}}|})
\end{aligned}
\]
where 
\begin{itemize}
\setlength{\itemsep}{3pt}
\setlength{\parsep}{0pt}
\setlength{\parskip}{0pt}
    \item $|Q_{\mathcal{P}}| = |Q| \times |Q_{ce}| \times |Q_{ec}| \times |Q_{sc}| \times |Q_{i}|$
    \item $|Q_{\mathcal{P}_{E}}| = |Q| \times |Q_{ce}| \times |Q_{ec}| \times |Q_{sc}| \times |Q_{i}| \times |Q_{s}|$
\end{itemize}
($|Q_{ce}| = 2^{|\Sigma_{c}|}$, $|Q_{ec}| = U+2$, $|Q_{sc}| = 2$, $|Q_{i}| \leq 2^{|Q|} + 1$)


 \subsection{Incremental synthesis: first $E$ and then $S$}
\label{subsec:Synthesis_E_S}

In this heuristic, we first synthesize the edit function $E$ to ensure the opacity, covertness and the marker-reachability of closed-loop system. Then, we synthesize the supervisor $S$ to ensure the safety and  nonblockingness. The details of the synthesis procedure are as follows: 

\vspace{0.1cm}
\noindent \textbf{Procedure 2:}
\begin{enumerate}[1.]
    \item Compute $\mathcal{P} = G||CE||EC||SC||I = (Q_{\mathcal{P}}, \Sigma_{\mathcal{P}}, \xi_{\mathcal{P}}, q_{\mathcal{P}}^{init},\\  Q_{\mathcal{P},m})$
    \item Generate $\mathcal{P}_{E}^{r} := (Q_{\mathcal{P}_{E}^{r}}, \Sigma_{\mathcal{P}_{E}^{r}}, \xi_{\mathcal{P}_{E}^{r}}, q_{\mathcal{P}_{E}^{r}}^{init}, Q_{\mathcal{P}_{E}^{r},m})$
    \begin{itemize}
    \setlength{\itemsep}{3pt}
    \setlength{\parsep}{0pt}
    \setlength{\parskip}{0pt}
        \item $Q_{\mathcal{P}_{E}^{r}} := Q_{\mathcal{P}} - Q_{5}$
        \begin{itemize}
        \setlength{\itemsep}{3pt}
        \setlength{\parsep}{0pt}
        \setlength{\parskip}{0pt}
            \item $Q_{5} := \{(q, q_{ce}, q_{ec}, q_{sc}, q_{i}) \in Q_{\mathcal{P}}|\, q_{i} = q^{unsafe} \vee q_{i} = \varnothing\}$
        \end{itemize}
        \item $\Sigma_{\mathcal{P}_{E}^{r}} := \Sigma_{\mathcal{P}}$
        \item $(\forall q, q' \in Q_{\mathcal{P}_{E}^{r}})(\forall \sigma \in \Sigma_{\mathcal{P}_{E}^{r}}) \, \xi_{\mathcal{P}}(q, \sigma) = q' \Leftrightarrow \xi_{\mathcal{P}_{E}^{r}}(q, \sigma) = q'$
        \item $q_{\mathcal{P}_{E}^{r}}^{init} := q_{\mathcal{P}}^{init}$
        \item $Q_{\mathcal{P}_{E}^{r},m} := Q_{\mathcal{P},m} - Q_{5}$
    \end{itemize}
    \item Synthesize a supervisor $E = (Q_{e}, \Sigma_{e}, \xi_{e}, q_{e}^{init}, Q_{e,m})$ over the edit function-control constraint $(\Sigma_{e,c}, \Sigma_{e,o})$ by treating $\mathcal{P}$ as the plant and $\mathcal{P}_{E}^{r}$ as the requirement such that $\mathcal{P}||E$ is marker-reachable and safe w.r.t. $\mathcal{P}_{E}^{r}$. If $E$ exists, go to Step 4; otherwise, end the procedure.
    \item Compute $\mathcal{P}_{S} = G||CE||EC||SC||I||E = (Q_{\mathcal{P}_{S}}, \Sigma_{\mathcal{P}_{S}}, \\ \xi_{\mathcal{P}_{S}}, q_{\mathcal{P}_{S}}^{init}, Q_{\mathcal{P}_{S},m})$.
    \item Generate $\mathcal{P}_{S}^{r} := (Q_{\mathcal{P}_{S}^{r}}, \Sigma_{\mathcal{P}_{S}^{r}}, \xi_{\mathcal{P}_{S}^{r}}, q_{\mathcal{P}_{S}^{r}}^{init}, Q_{\mathcal{P}_{S}^{r},m})$ 
    \begin{itemize}
    \setlength{\itemsep}{3pt}
    \setlength{\parsep}{0pt}
    \setlength{\parskip}{0pt}
        \item $Q_{\mathcal{P}_{S}^{r}} := Q_{\mathcal{P}_{S}} - Q_{6}$
        \begin{itemize}
        \setlength{\itemsep}{3pt}
        \setlength{\parsep}{0pt}
        \setlength{\parskip}{0pt}
            \item $Q_{6} := \{(q, q_{ce}, q_{ec}, q_{sc}, q_{i}, q_{e}) \in Q_{\mathcal{P}_{S}}|\, q \in Q_{avoid}\}$
        \end{itemize}
        \item $\Sigma_{\mathcal{P}_{S}^{r}} := \Sigma_{\mathcal{P}_{S}}$
        \item $(\forall q, q^{'} \in Q_{\mathcal{P}_{S}^{r}})(\forall \sigma \in \Sigma_{\mathcal{P}_{S}^{r}}) \, \xi_{\mathcal{P}_{S}}(q, \sigma) = q^{'} \Leftrightarrow \xi_{\mathcal{P}_{S}^{r}}(q, \sigma) = q^{'}$
        \item $q_{\mathcal{P}_{S}^{r}}^{init} := q_{\mathcal{P}_{S}}^{init}$
        \item $Q_{\mathcal{P}_{S}^{r},m} := Q_{\mathcal{P}_{S},m} - Q_{6}$
    \end{itemize}
    \item Synthesize a supervisor $S = (Q_{s}, \Sigma_{s}, \xi_{s}, q_{s}^{init}, Q_{s,m})$ over the supervisor-control constraint $(\Sigma_{s,c}, \Sigma_{s,o})$ by treating $\mathcal{P}_{S}$ as the plant and $\mathcal{P}_{S}^{r}$ as the requirement such that $\mathcal{P}_{S}||S$ is nonblocking and safe w.r.t. $\mathcal{P}_{S}^{r}$.
\end{enumerate}

In the above procedure, Steps 1-3 focus on the synthesis of the edit function $E$ and Steps 4-6 focus on the synthesis of the supervisor $S$. In the part regarding the synthesis of edit function, $\mathcal{P}$ and $\mathcal{P}_{E}^{r}$ are treated as the plant and the requirement, respectively. The requirement $\mathcal{P}_{E}^{r}$ is generated from $\mathcal{P}$ by removing the states where the intruder reaches the state $q^{unsafe}$ or $\varnothing$. Then, at Step 3, we compute the edit function $E$ that can ensure the safety w.r.t. $\mathcal{P}_{E}^{r}$ and the marker-reachability. At this step, the nonblockingness of the closed-loop system  is hard to ensure since the sending of control commands by the supervisor is uncontrollable to the edit function.
In the part regarding the synthesis of supervisor, $\mathcal{P}_{S}$ and $\mathcal{P}_{S}^{r}$ are treated as the plant and the requirement, respectively. The plant $\mathcal{P}_{S}$ is generated based on the edit function $E$ synthesized in Steps 1-3. At Step 5, the requirement $\mathcal{P}_{S}^{r}$ is generated from $\mathcal{P}_{S}$ by removing the set of states $Q_{6}$, which is not allowed by the user requirement.
Finally, we compute the supervisor $S$ that can ensure the safety w.r.t. $\mathcal{P}_{S}^{r}$ and nonblockingness at Step 6.  \textbf{Procedure 2} clearly terminates within finite steps. 

\vspace{0.1cm}

\emph{Theorem IV.3:} Given any plant $G$, command execution component $CE$, edit constraints $EC$, supervisor constraints $SC$, and intruder $I$, the computed $E$ in \textbf{Procedure 2}, if not empty, with any supervisor $\widetilde{S} = (\widetilde{Q_{s}}, \widetilde{\Sigma_{s}}, \widetilde{\xi_{s}}, \widetilde{q_{s}^{init}}, \widetilde{Q_{s,m}})$ is an opaque and covert edit function-supervisor pair w.r.t. the edit function-control constraint $(\Sigma_{e,c}, \Sigma_{e,o})$ and supervisor-control constraint $(\Sigma_{s,c}, \Sigma_{s,o})$ for $G$, $CE$, $EC$, $SC$, and $I$.


\emph{Proof:} Based on Step 2 in \textbf{Procedure 2}, the set of states $Q_{5}$ has been removed in the requirement $Q_{\mathcal{P}_{E}^{r}}$, i.e., they are treated as ``bad'' states in the synthesis of $E$. Thus, any state in $\{(q, q_{ce}, q_{ec}, q_{sc}, q_{i}, q_{e}) \in Q \times Q_{ce} \times Q_{ec} \times Q_{sc} \times Q_{i} \times Q_{e}|\, q_{i} = q^{unsafe} \vee q_{i} = \varnothing\}$ is not reachable in $\mathcal{P}||E$, which means that any state in $\{(q, q_{ce}, q_{ec}, q_{sc}, q_{i}, q_{e}, q_{s}) \in Q \times Q_{ce} \times Q_{ec} \times Q_{sc} \times Q_{i} \times Q_{e} \times \widetilde{Q_{s}}|\, q_{i} = q^{unsafe} \vee q_{i} = \varnothing\}$ is not reachable in $\mathcal{P}||E||\widetilde{S}$. Based on the definition IV.1 and definition IV.2, the proof is completed. \hfill $\blacksquare$

\vspace{0.1cm}



\vspace{0.1cm}

\emph{Theorem IV.4:} Given any plant $G$, command execution component $CE$, edit constraints $EC$, supervisor constraints $SC$, and intruder $I$, the computed $E$ and $S$ in \textbf{Procedure 2}, if not empty, could satisfy the following goals:
\begin{itemize}
\setlength{\itemsep}{3pt}
\setlength{\parsep}{0pt}
\setlength{\parskip}{0pt}
    \item $G||CE||EC||SC||I||E||S$ is nonblocking and any state in $\{(q, q_{ce}, q_{ec}, q_{sc}, q_{i}, q_{e}, q_{s}) \in Q \times Q_{ce} \times Q_{ec} \times Q_{sc} \times Q_{i} \times Q_{e} \times Q_{s}|\, q \in Q_{avoid}\}$ is not reachable in $G||CE||EC||SC||I||E||S$.
    \item $E$ combined with $S$ is an opaque edit function-supervisor pair w.r.t. the edit function-control constraint $(\Sigma_{e,c}, \Sigma_{e,o})$ and supervisor-control constraint $(\Sigma_{s,c}, \Sigma_{s,o})$ for $G$, $CE$, $EC$, $SC$, and $I$.
    \item $E$ combined with $S$ is a covert edit function-supervisor pair w.r.t. the edit function-control constraint $(\Sigma_{e,c}, \Sigma_{e,o})$ and supervisor-control constraint $(\Sigma_{s,c}, \Sigma_{s,o})$ for $G$, $CE$, $EC$, $SC$, and $I$.
\end{itemize}

\emph{Proof:} Based on Step 6 of \textbf{Procedure 2}, the synthesized $S$ should satisfy that $\mathcal{P}_{S}||S$ is nonblocking, that is, $G||CE||EC||SC||I||E||S$ is nonblocking. In addition, at Step 5, the set of states $Q_{6}$ has been removed from $\mathcal{P}_{S}^{r}$, i.e., they are treated as ``bad'' states in the synthesis of $S$. Thus, any state in $\{(q, q_{ce}, q_{ec}, q_{sc}, q_{i}, q_{e}, q_{s}) \in Q \times Q_{ce} \times Q_{ec} \times Q_{sc} \times Q_{i} \times Q_{e} \times Q_{s}|\, q \in Q_{avoid}\}$ is not reachable in $\mathcal{P}_{S}||S = G||CE||EC||SC||I||E||S$.
Based on Theorem IV.3, the computed $E$ in \textbf{Procedure 2} with any supervisor is an opaque and covert edit function-supervisor pair, thus, the computed $E$ with the computed $S$ in \textbf{Procedure 2} is also an opaque and covert edit function-supervisor pair. This completes the proof. \hfill $\blacksquare$

Next, we shall analyze the computational complexity of \textbf{Procedure 2}.
By adopting the normality based synthesis approach in \cite{WangStateControl2018}, the complexity is 
\[
O(|\Sigma_{\mathcal{P}}|2^{|Q_{\mathcal{P}}|} + |\Sigma_{\mathcal{P}}||Q_{\mathcal{P}_{S}}|^{2}4^{|Q_{\mathcal{P}_{S}}|})= O(|\Sigma_{\mathcal{P}}||Q_{\mathcal{P}_{S}}|^{2}4^{|Q_{\mathcal{P}_{S}}|})
\]
where 
\begin{itemize}
\setlength{\itemsep}{3pt}
\setlength{\parsep}{0pt}
\setlength{\parskip}{0pt}
    \item $|Q_{\mathcal{P}}| = |Q| \times |Q_{ce}| \times |Q_{ec}| \times |Q_{sc}| \times |Q_{i}|$
    \item $|Q_{\mathcal{P}_{S}}| = |Q| \times |Q_{ce}| \times |Q_{ec}| \times |Q_{sc}| \times |Q_{i}| \times |Q_{e}|$
\end{itemize}

\section{Example}
\label{sec:example}
In this section, we shall present an example to show the effectiveness of the proposed method to synthesize the edit function and the supervisor for opacity enforcement in the supervisory control of discrete-event systems.

\begin{figure}[htbp]
\begin{center}
\includegraphics[height=5.7cm]{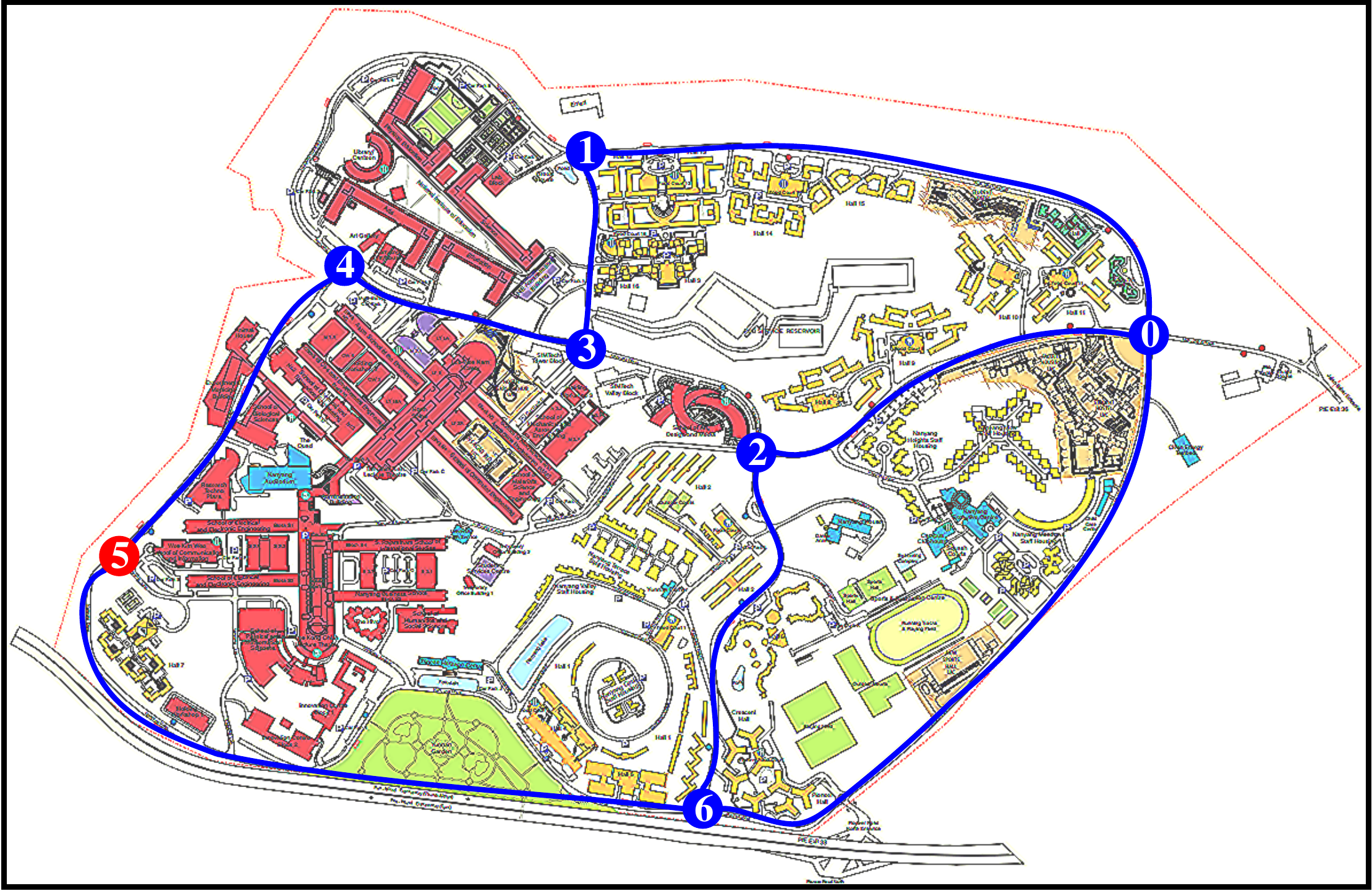}   
\caption{The Nanyang Technological University campus map}
\label{fig:The Nanyang Technological University campus map}
\end{center}        
\end{figure}

\textbf{Example 5.1} We adapt the location-based privacy example of \cite{Wu2014LocationPrivacy} for an illustration. In this example, a batch of confidential experiment devices are transported by an autonomous vehicle to the EEE building of the Nanyang Technological University. After completing the transportation task, the vehicle is required to leave the campus.
The location of the vehicle is obtained based on the Global Positioning System (GPS) and the location information acquisition channel is eavesdropped by the intruder whose target is to infer whether the confidential experiment devices have been transported to the EEE building. The Nanyang Technological University campus map is shown in Fig. \ref{fig:The Nanyang Technological University campus map}, where we discretize the model by selecting seven locations as states, marked by blue and red circles, and several connection routes between those locations, marked by blue lines. Location (state) 5 represents the EEE building, which is the secret location (state) that the intruder intends to infer. 

In this example, $\Sigma = \{a, b, c, b_{uc}, c_{uo}\}$, where $a$, $b$, and $c$ represent turning right, going straight, and turning left. $b_{uc}$ also represents going straight but is uncontrollable. $c_{uo}$ also represents turning left but is unobservable since there exists shade of trees on some route, resulting in that GPS service is not available. Thus,  $\Sigma_{uc} = \{b_{uc}\}$ and $\Sigma_{uo} = \{c_{uo}\}$. $\Sigma_{o,I} = \Sigma_{s,E} = \Sigma_{o,E} = \Sigma_{o} = \{a, b, c, b_{uc}\}$. $U = 1$. 
Since $b^{\#}$ and $b_{uc}^{\#}$ essentially represent the same event, going straight, for the intruder and supervisor, we have $\Sigma_{s,E}^{\#} = \{a^{\#}, b^{\#}, c^{\#}\}$.
Since enabling $c$ and enabling $c_{uo}$ by the supervisor essentially represent the same control decision, they should exist together in any control command. Thus, $\Gamma = \{v_{1}, v_{2}, v_{3}, v_{4}, v_{5}, v_{6}, v_{7}\}$, where $v_{1} = \{a\}$, $v_{2} = \{b\}$, $v_{3} = \{c, c_{uo}\}$, $v_{4} = \{a, b\}$, $v_{5} = \{a, c, c_{uo}\}$, $v_{6} = \{b, c, c_{uo}\}$, and $v_{7} = \{a, b, c, c_{uo}\}$.

\begin{figure}[htbp]
\begin{center}
\includegraphics[height=3.4cm]{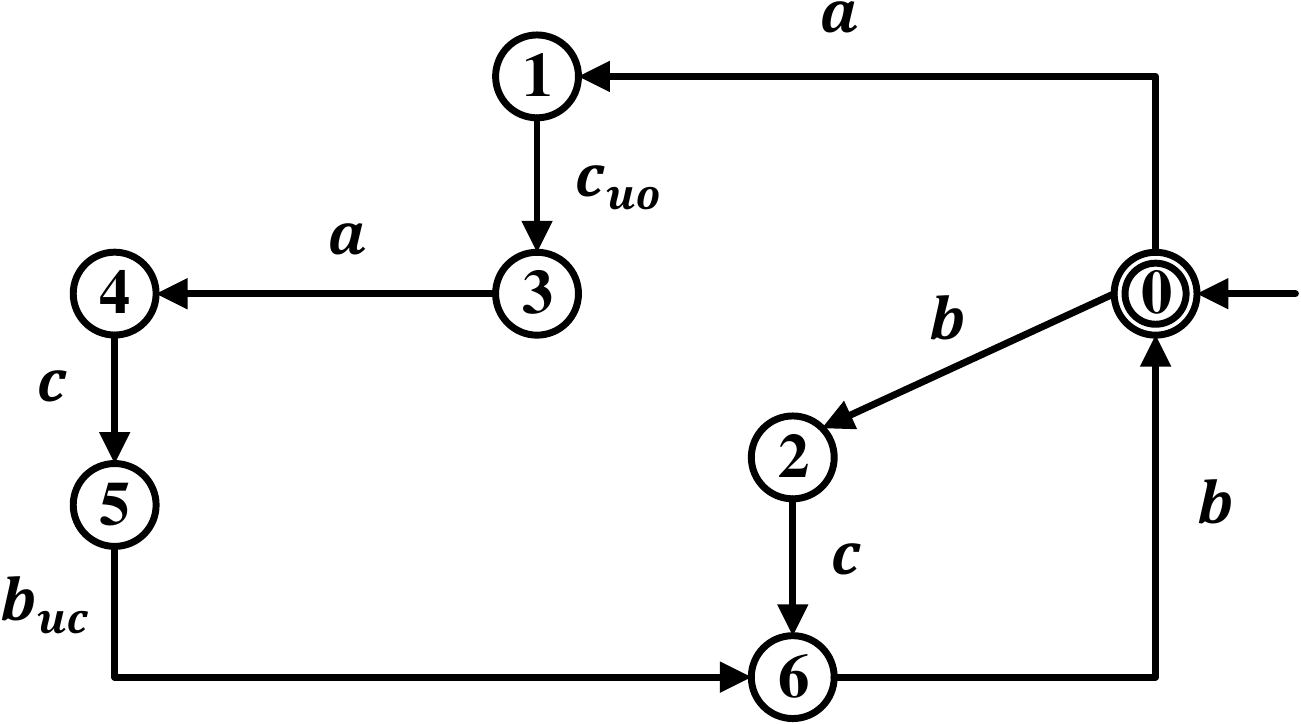}   
\caption{Plant $G$}
\label{fig:Plant G}
\end{center}        
\end{figure}

\begin{figure}[htbp]
\begin{center}
\includegraphics[height=3.4cm]{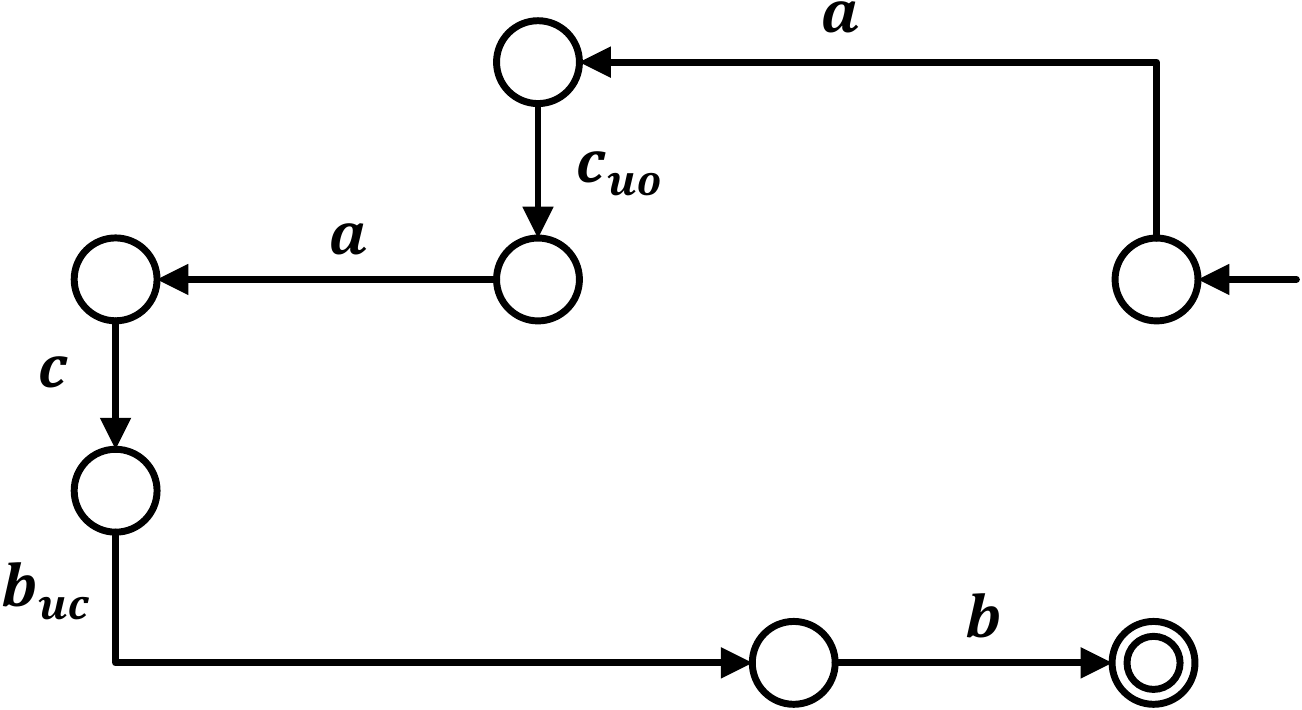}   
\caption{Requirement of the plant $G$}
\label{fig:Specification K}
\end{center}        
\end{figure}

\begin{figure}[htbp]
\begin{center}
\includegraphics[height=6.7cm]{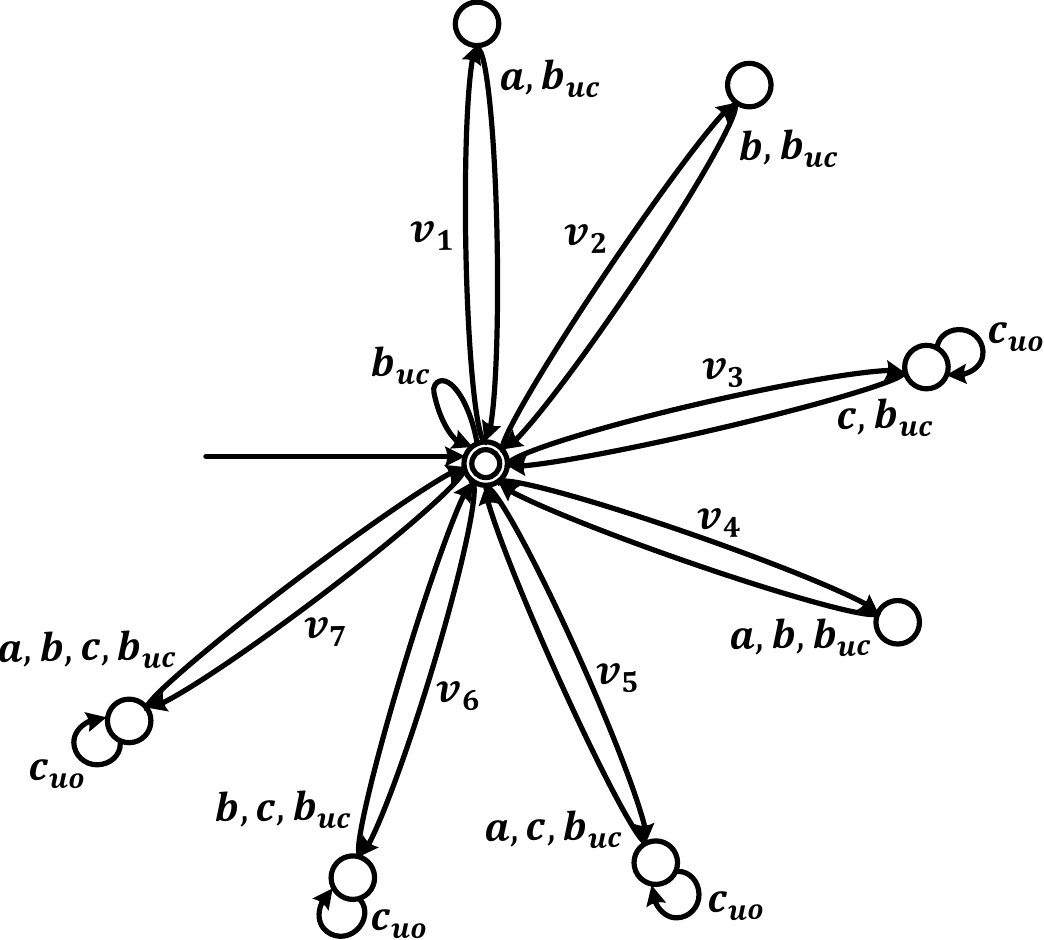}   
\caption{Command execution $CE$}
\label{fig:Command execution CE}
\end{center}        
\end{figure}

\begin{figure}[htbp]
\begin{center}
\includegraphics[height=2.7cm]{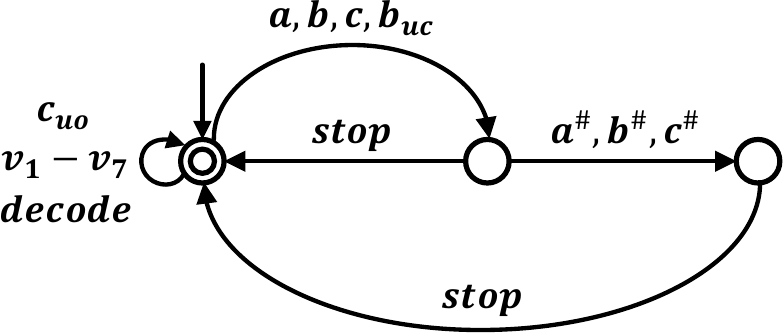}   
\caption{Edit constraints $EC$}
\label{fig:Edit constraints EC}
\end{center}        
\end{figure}

\begin{figure}[htbp]
\begin{center}
\includegraphics[height=3cm]{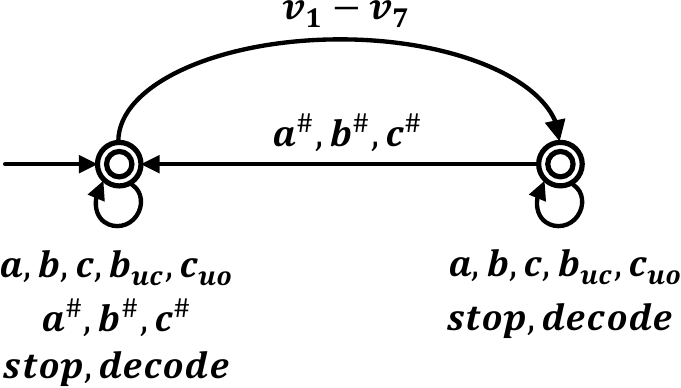}   
\caption{Supervisor constraints $SC$}
\label{fig:Supervisor constraints SC}
\end{center}        
\end{figure}

\begin{figure}[htbp]
\begin{center}
\includegraphics[height=4.3cm]{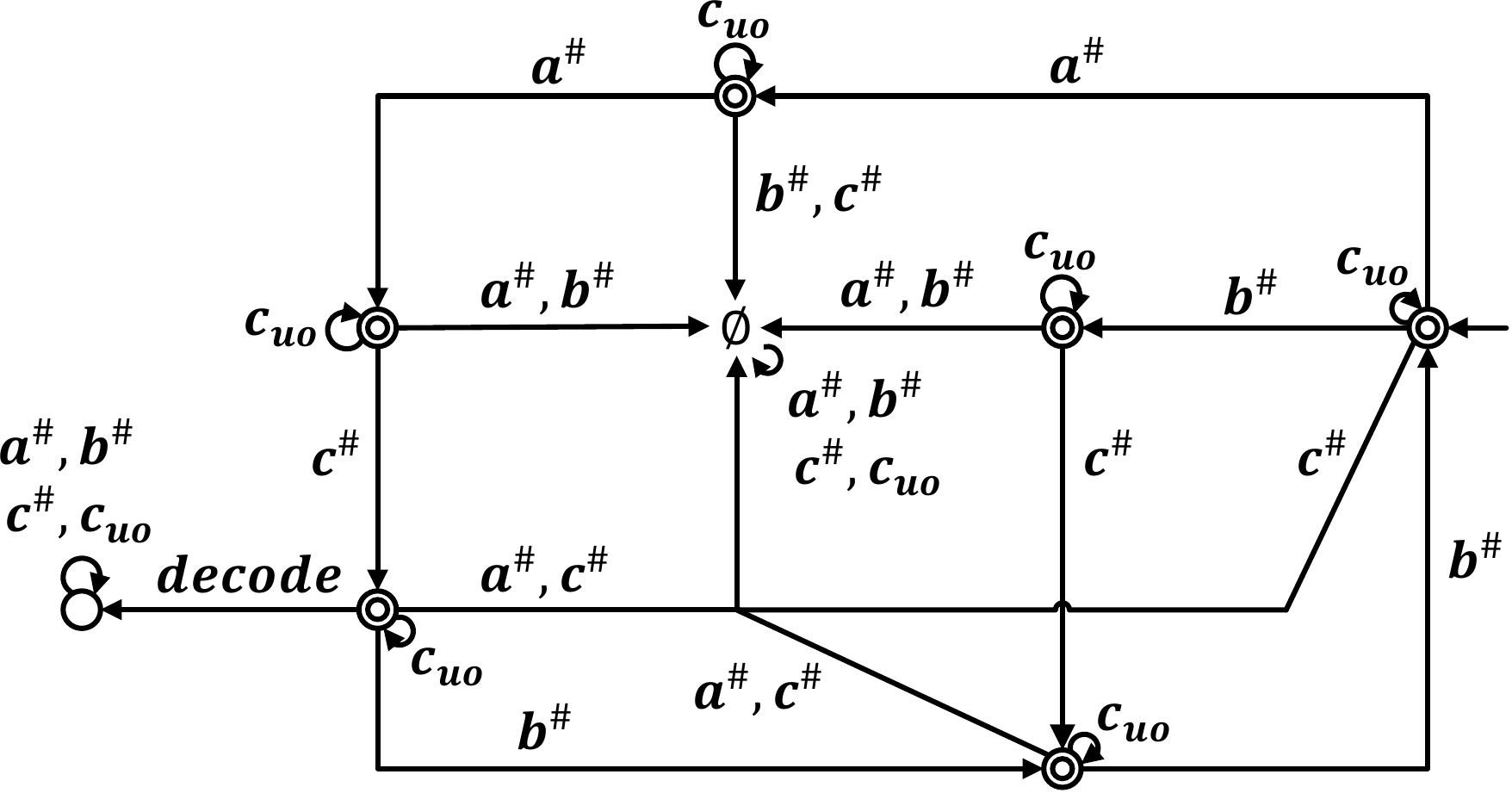}   
\caption{Intruder $I$}
\label{fig:Intruder I}
\end{center}        
\end{figure}

Plant $G$ is shown in Fig. \ref{fig:Plant G}, where the state 5 is the secret state. The requirement of $G$ is shown in Fig. \ref{fig:Specification K}. 
Command execution automaton $CE$ is shown in Fig. \ref{fig:Command execution CE}. Edit constraints $EC$ is shown in Fig. \ref{fig:Edit constraints EC}. Supervisor constraints $SC$ is shown in Fig. \ref{fig:Supervisor constraints SC}. Intruder $I$ is shown in Fig. \ref{fig:Intruder I}.

\begin{figure*}[htbp]
\begin{center}
\includegraphics[height=12.9cm]{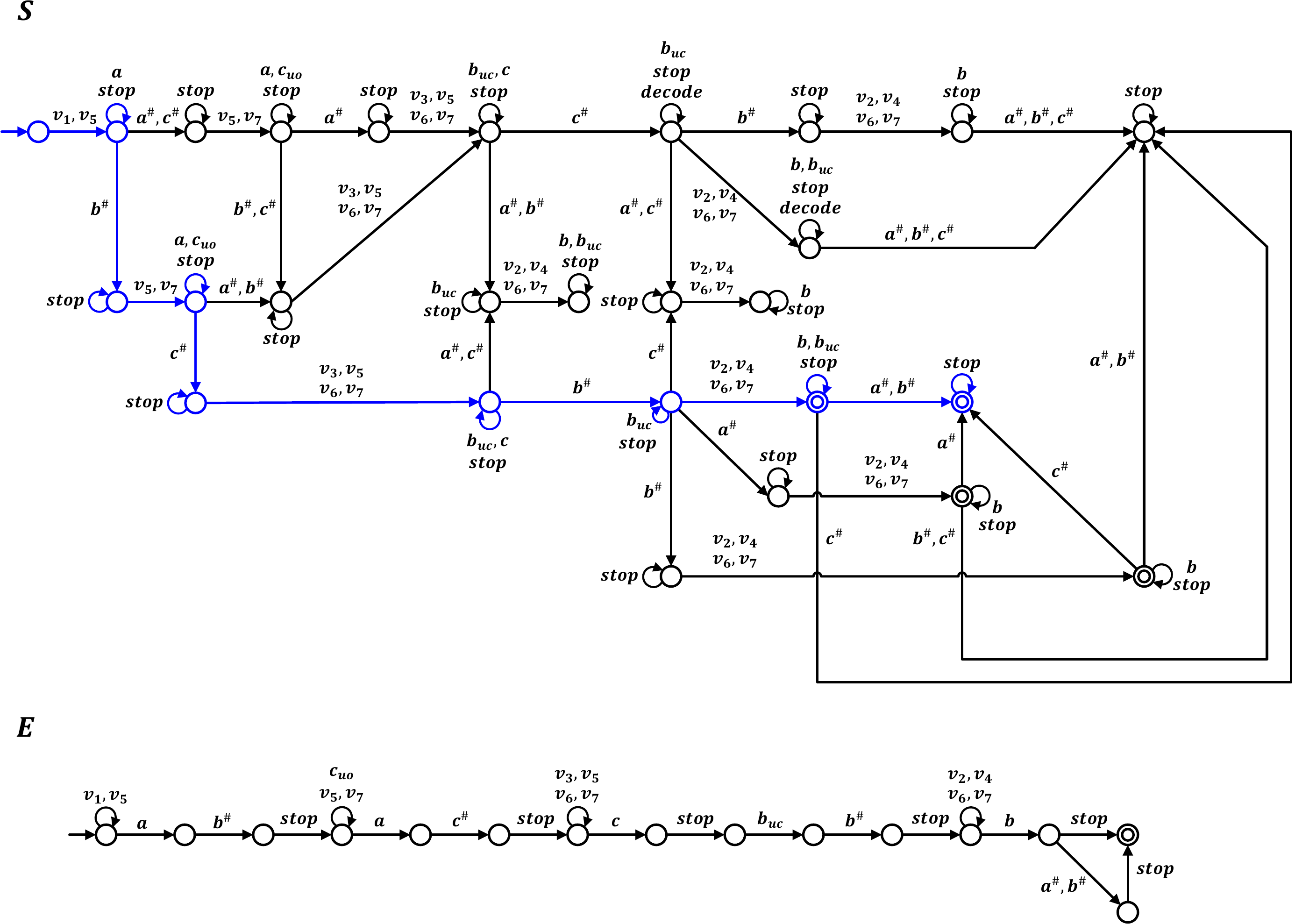}   
\caption{Synthesized $S$ and $E$ by adopting incremental synthesis from $S$ to $E$}
\label{fig:Synthesized S_E}
\end{center}        
\end{figure*}

\begin{figure*}[htbp]
\begin{center}
\includegraphics[height=13cm]{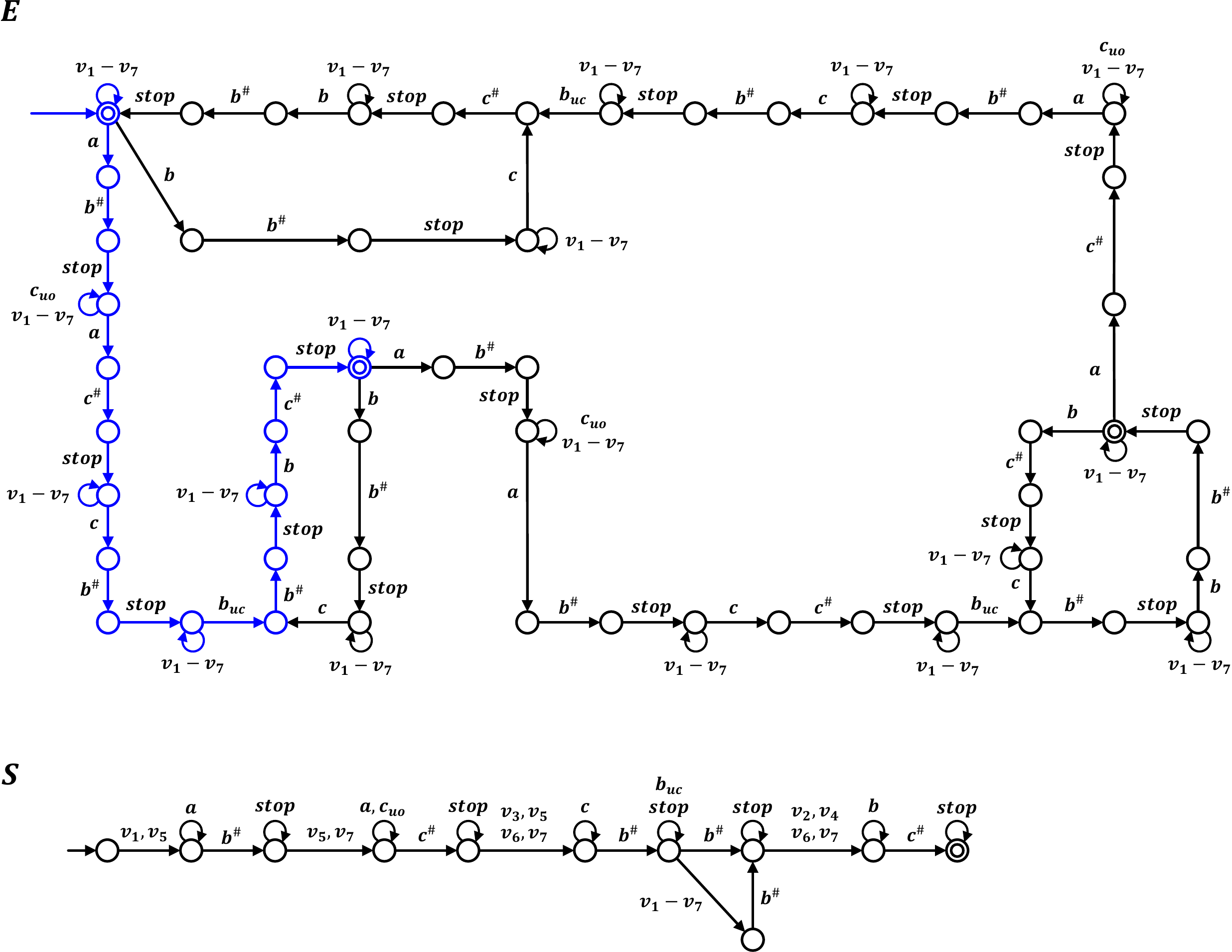}   
\caption{Synthesized $E$ and $S$ by adopting incremental synthesis from $E$ to $S$}
\label{fig:Synthesized E_S}
\end{center}        
\end{figure*}

We use \textbf{SuSyNA} \cite{SuSyNA} to synthesize the edit function and the supervisor based on the procedures proposed in Section \ref{subsec:Synthesis_S_E} and \ref{subsec:Synthesis_E_S}.
By adopting the incremental synthesis from $S$ to $E$, the synthesized supervisor and edit function are shown in Fig. \ref{fig:Synthesized S_E}. By adopting the incremental synthesis from $E$ to $S$, the synthesized edit function and supervisor are shown in Fig. \ref{fig:Synthesized E_S}. In carrying out the incremental synthesis from $E$ to $S$, we slightly restrict the capabilities of the edit function because the synthesized $E$ at the first step might always delete the events in $\Sigma_{o}$ since $\Sigma_{o} = \Sigma_{o,E} = \Sigma_{s,E}$ in this example, resulting in the situation that the supervisor would not observe any information and then does not issue any control command. 
Thus, in this example, we shall implement the incremental synthesis from $E$ to $S$ by assuming that the edit function cannot delete editable events. In addition, since $U=1$, the edit function can only replace editable events.



It can be checked that both of the two synthesized results can achieve the following goals: 1) the closed-loop system  is nonblocking and satisfy the  requirement (see Fig.~\ref{fig:Specification K}); 2) the intruder could never infer that plant $G$ has reached the secret state; 3) the edit function always remains covert. Next, we shall present some explanations for the two synthesized results. In Fig. \ref{fig:Synthesized S_E} (Fig. \ref{fig:Synthesized E_S}), under the cooperation of the synthesized $E$ and $S$, the actual strategies adopted by $S$ ($E$) are marked as the blue highlighted parts. 

The intuitive explanation of the two synthesized results is that: the edit function $E$ would always alter the authentic sensor readings into different values to trick the intruder $I$ such that $I$ believes that the motion path of the autonomous vehicle is $bcb$ (recall that the intruder does not know the  specification), a path in the campus map that would not expose the secret state; meanwhile, based on the changed sensor readings, the supervisor $S$ would issue the appropriate control command to guarantee that the true motion path of the vehicle is $ac_{uo}acb_{uc}b$, a path that could fulfill the specification $K$.

The details of the synthesized strategies in Fig. \ref{fig:Synthesized S_E} and \ref{fig:Synthesized E_S} are as follows: 
In the first few steps, the strategies of the synthesized edit functions combined with supervisors by two incremental synthesis methods are the same. At the initial state, the supervisor $S$ issues the initial control command $v_{1}$ or $v_{5}$. After receiving the control command $v_{1}$ or $v_{5}$, plant $G$ would execute event $a$, which could be observed by the edit function $E$. Then $E$ changes $a$ to $b^{\#}$, triggering $S$ to issue the control command $v_{5}$ or $v_{7}$, after which plant $G$ would execute event $c_{uo}$. Since $c_{uo} \in \Sigma_{uo}$, $v_{5}$ or $v_{7}$ would be reused by plant $G$ and event $a$ is then executed. After observing $a$, $E$ would change it to $c^{\#}$, triggering $S$ to issue the control command $v_{3}$ or $v_{5}$ or $v_{6}$ or $v_{7}$. Then the event $c$ is executed by plant $G$. Afterwards, the strategies of the synthesized edit functions combined with supervisors by two incremental synthesis methods are different:
\begin{enumerate}[1.]
    \item Incremental synthesis from $S$ to $E$: After observing $c$, $E$ would delete it, resulting in that $S$ would not issue any control command. $E$ would wait until the uncontrollable event $b_{uc}$ is fired by plant $G$, then it would change $b_{uc}$ to $b^{\#}$, triggering $S$ to issue the control command $v_{2}$ or $v_{4}$ or $v_{6}$ or $v_{7}$. Then event $b$ is executed by plant $G$, after which $E$ could either delete $b$ or change $b$ to anyone of $a^{\#}$ and $b^{\#}$. In this case, what the intruder observes during the whole process is $b^{\#}c^{\#}b^{\#}$ or $b^{\#}c^{\#}b^{\#}a^{\#}$ or $b^{\#}c^{\#}b^{\#}b^{\#}$, anyone of which would not break the opacity and covertness property.
    \item Incremental synthesis from $E$ to $S$: After observing $c$, $E$ would change it to $b^{\#}$, which could be observed by $S$. Then, two situations might happen: 
    \begin{enumerate}[a.]
        \item The uncontrollable event $b_{uc}$ is fired immediately after $E$ replaces $c$ with $b^{\#}$, which preempts the event of issuing a control command by $S$. After observing $b_{uc}$, $E$ would replace it with $b^{\#}$, resulting in that $S$ would observe $b^{\#}$ again and issue the control command $v_{2}$ or $v_{4}$ or $v_{6}$ or $v_{7}$;
        \item The uncontrollable event $b_{uc}$ is not fired immediately after $E$ replaces $c$ with $b^{\#}$. Then $S$ issues any control command from $v_{1}$ to $v_{7}$, after which the event $b_{uc}$ is executed. The observation of $b_{uc}$ would trigger $E$ to replace it with $b^{\#}$, resulting in that $S$ would issue the control command $v_{2}$ or $v_{4}$ or $v_{6}$ or $v_{7}$.
    \end{enumerate}
    Then event $b$ is executed by plant $G$, after which $E$ would change it to $c^{\#}$. In this case, what the intruder observes during the whole process is $b^{\#}c^{\#}b^{\#}b^{\#}c^{\#}$, which would not break the opacity and covertness property.
\end{enumerate}

To illustrate the advantage of our proposed incremental synthesis method, we also adopt the aggregative synthesis based approach proposed in \cite{Su2010Aggregative} to synthesize the desired edit function and supervisor for this example. By directly using the {\bf make\_supervisor} operation in \textbf{SuSyNA}, we always generate an empty distributed supervisor because the approach in \cite{Su2010Aggregative} would always try to find a nonblocking local supervisor at each synthesis step, resulting in an empty solution even at the first synthesis step, no matter whether $S$ or $E$ is synthesized first. 

\section{Conclusions}
\label{sec:conclusions}
In this paper, we propose a novel privacy-preserving supervisory control problem, and present heuristic approaches to co-synthesize edit function and supervisor for opacity enforcement and requirement satisfaction in discrete-event systems. By modeling the co-synthesis problem as a distributed supervisor synthesis problem in the Ramadge-Wonham supervisory control framework, our solutions allow existing synthesis tools such as SuSyNA, Supremica, or TCT to be used. 
In our future works, we shall explore other
powerful distributed synthesis heuristics to generate better distributed solutions for the privacy-preserving control problem. 



\begin{IEEEbiography}[
{
\includegraphics[width=1.0in,height=1.40in,clip,keepaspectratio]{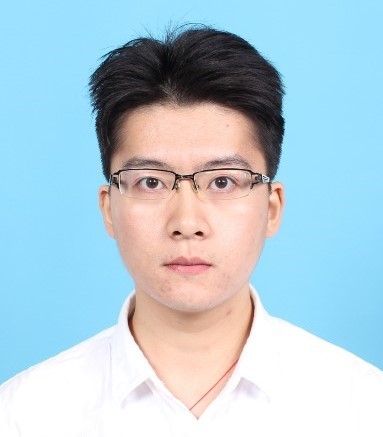}
}
]
{Ruochen Tai}
received the B.E. degree in electrical engineering from the Nanjing University of Science and Technology in 2016, and M.S. degree in automaton from the Shanghai Jiao Tong University in 2019. He is currently pursuing the Ph.D. degree with Nanyang Technological University, Singapore. His current research interests include security issue of cyber-physical systems, multi-robot systems, safe autonomy in cyber-physical-human systems, formal methods, and discrete-event systems.
\end{IEEEbiography}
\begin{IEEEbiography}[
{
\includegraphics[width=1.0in,height=1.40in,clip,keepaspectratio]{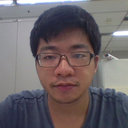}
}
]
{Liyong Lin}
received the B.E. degree and Ph.D. degree in electrical engineering in 2011 and 2016, respectively, both from Nanyang Technological University, where he has also worked as a project officer. From June 2016 to October 2017, he was a postdoctoral fellow at the University of Toronto. Since December 2017, he has been working as a research fellow at the Nanyang Technological University. His main research interests include supervisory control theory, formal methods and machine learning. He previously was an intern in the Data Storage Institute, Singapore, where he worked on single and dual-stage servomechanism of hard disk drives.
\end{IEEEbiography}
\begin{IEEEbiography}[
{
\includegraphics[width=1.0in,height=1.20in,clip,keepaspectratio]{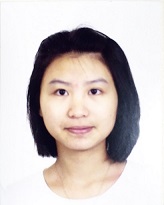}
}
]
{Yuting Zhu}
received the B.S. degree from Southeast University, Jiangsu, China, in 2016. She is currently pursuing the Ph.D. degree with Nanyang Technological University, Singapore. Her research interests include networked control and cyber security of discrete event systems.
\end{IEEEbiography}
\begin{IEEEbiography}[
{
\includegraphics[width=1in,height=1.3in,clip,keepaspectratio]{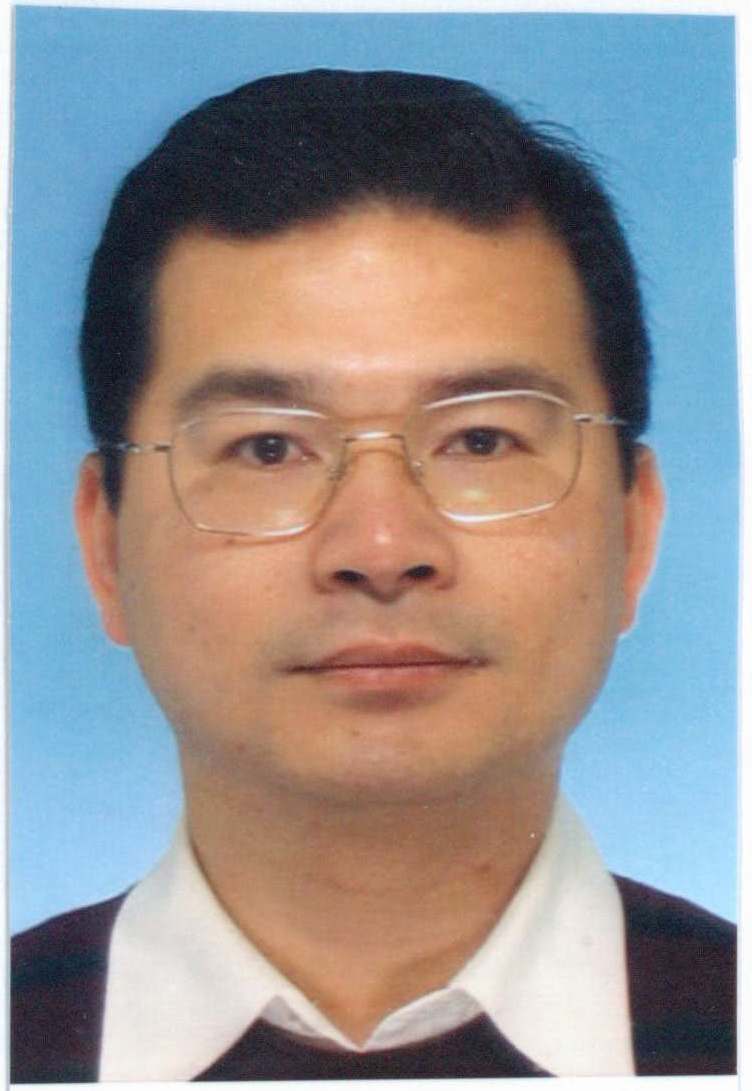}
}
]
{Rong Su} received the Bachelor of Engineering degree from University of Science and Technology of China in 1997, and the Master of Applied Science degree and PhD degree from University of Toronto, in 2000 and 2004, respectively. He was affiliated with University of Waterloo and Technical University of Eindhoven before he joined Nanyang Technological University in 2010. Currently, he is an associate professor in the School of Electrical and Electronic Engineering. Dr. Su's research interests include multi-agent systems, cyber security of discrete-event systems, supervisory control, model-based fault diagnosis, control and optimization in complex networked systems with applications in flexible manufacturing, intelligent transportation, human-robot interface, power management and green buildings. In the aforementioned areas he has more than 220 journal and conference publications, and 5 granted USA/Singapore patents. Dr. Su is a senior member of IEEE, and an associate editor for Automatica, Journal of Discrete Event Dynamic Systems: Theory and Applications, and Journal of Control and Decision. He was the chair of the Technical Committee on Smart Cities in the IEEE Control Systems Society in 2016-2019, and is currently the chair of IEEE Control Systems Chapter, Singapore, and a co-chair of IEEE Robotic and Automation Society Technical Committee on Automation in Logistics.

\end{IEEEbiography}

\end{document}